\documentclass[twocolumn]{aastex631}
\usepackage{upgreek}
\usepackage{amsmath}

\newcommand\target{Gaia20eae}
\newcommand\konkoly{Konkoly Observatory, Research Centre for Astronomy and Earth Sciences,\\Eötvös Loránd Research Network (ELKH), Konkoly-Thege Mikl\'os \'ut 15--17, 1121 Budapest, Hungary}
\newcommand\elte{ELTE E\"otv\"os Lor\'and University, Institute of Physics, P\'azm\'any P\'eter s\'et\'any 1/A, 1117 Budapest, Hungary}
\newcommand\roma{INAF-Osservatorio Astronomico di Roma, Via di Frascati 33, 00078 Monte Porzio Catone, Italy}

\newcommand{\lsun}{\mbox{L}_\odot}
\newcommand{\rsun}{\mbox{R}_\odot}
\newcommand{\msun}{\mbox{M}_\odot}

\newcommand{\lstar}{L_\star}
\newcommand{\mstar}{M_\star}
\newcommand{\rstar}{R_\star}

\shorttitle{New EXor candidate: Gaia20eae}
\shortauthors{Cruz-Saénz de Miera et al.}

\begin{document}
\title{Recurrent strong outbursts of an EXor-like young eruptive star Gaia20eae}

\correspondingauthor{Fernando Cruz-Sáenz de Miera}
\email{cruzsaenz.fernando@csfk.org}

\author[0000-0002-4283-2185]{Fernando Cruz-Sáenz de Miera}
\affiliation{\konkoly{}}

\author[0000-0001-7157-6275]{Ágnes Kóspál}
\affiliation{\konkoly{}}
\affiliation{Max Planck Institute for Astronomy, Königstuhl 17, 69117 Heidelberg, Germany}
\affiliation{\elte{}}

\author[0000-0001-6015-646X]{Péter Ábrahám}
\affiliation{\konkoly{}}
\affiliation{\elte{}}

\author[0000-0003-4099-1171]{Sunkyung Park}
\affiliation{\konkoly{}}

\author[0000-0002-3632-1194]{Zsófia Nagy}
\affiliation{\konkoly{}}

\author[0000-0001-5018-3560]{Michał Siwak}
\affiliation{\konkoly{}}

\author[0000-0002-7538-5166]{Mária Kun}
\affiliation{\konkoly{}}

\author[0000-0002-5261-6216]{Eleonora Fiorellino}
\affiliation{\konkoly{}}
\affiliation{\roma{}}

\author[0000-0001-9830-3509]{Zsófia Marianna Szabó}
\affiliation{\konkoly{}}
\affiliation{Max-Planck-Institut für Radioastonomie, Auf dem Hügel 69, 53121, Bonn, Germany}
\affiliation{Scottish Universities Physics Alliance (SUPA), School of Physics and Astronomy,\\University of St Andrews, North Haugh, St Andrews, KY16 9SS, UK}

\author[0000-0002-0666-3847]{Simone Antoniucci}
\affiliation{\roma{}}

\author{Teresa Giannini}
\affiliation{\roma{}}

\author[0000-0002-9190-0113]{Brunella Nisini}
\affiliation{\roma{}}

\author[0000-0002-2046-4131]{László Szabados}
\affiliation{\konkoly{}}


\author[0000-0002-1792-546X]{Levente Kriskovics}
\affiliation{\konkoly{}}

\author{Andr\'as Ordasi}
\affiliation{\konkoly{}}

\author[0000-0002-1698-605X]{R\'obert Szak\'ats}
\affiliation{\konkoly{}}

\author[0000-0002-6471-8607]{Kriszti\'an Vida}
\affiliation{\konkoly{}}

\author[0000-0001-8764-7832]{J\'ozsef Vink\'o}
\affiliation{\konkoly{}}

\author[0000-0001-6434-9429]{Pawe{\l} Zieli{\'n}ski}
\affiliation{Astronomical Observatory, University of Warsaw, Al. Ujazdowskie 4, 00-478, Warsaw, Poland}
\affiliation{Institute of Astronomy, Faculty of Physics, Astronomy and Informatics,\\Nicolaus Copernicus University in Toru{\'n}, ul. Grudzi{\,a}dzka 5, 87-100 Toru{\'n}, Poland}

\author[0000-0002-9658-6151]{{\L}ukasz Wyrzykowski}
\affiliation{Astronomical Observatory, University of Warsaw, Al. Ujazdowskie 4, 00-478, Warsaw, Poland}

\author{David García-Álvarez}
\affiliation{Instituto de Astrofísica de Canarias, Avenida Vía Láctea, E-38205 La Laguna, Tenerife, Spain}
\affiliation{Grantecan S.A., Centro de Astrofísica de La Palma, Cuesta de San José, E-38712 Breña Baja, La Palma, Spain}

\author{Marek Dr{\'o}{\.z}d{\.z}}
\affiliation{Mount Suhora Astronomical Observatory, Cracow Pedagogical University, ul. Podchorazych 2, 30-084 Krak{\'o}w, Poland}

\author[0000-0002-6293-9940]{Waldemar Og{\l}oza}
\affiliation{Mount Suhora Astronomical Observatory, Cracow Pedagogical University, ul. Podchorazych 2, 30-084 Krak{\'o}w, Poland}

\author{Eda Sonbas}
\affiliation{Department of Physics, Adiyaman University, 02040 Adiyaman, Turkey}

\begin{abstract}
We present follow-up photometric and spectroscopic observations, and subsequent analysis of Gaia20eae.
This source triggered photometric alerts during 2020 after showing a $\sim$3\,mag increase in its brightness.
Its Gaia Alert light curve showed the shape of a typical eruptive young star.
We carried out observations to confirm Gaia20eae as an eruptive young star and classify it.
Its pre-outburst spectral energy distribution shows that Gaia20eae is a moderately embedded Class~II object with $L_\mathrm{bol} = 7.22$\,L$_\odot$.
The color-color and color-magnitude diagrams indicate that the evolution in the light curve is mostly gray.
Multiple epochs of the H$\upalpha$ line profile suggest an evolution of the accretion rate and winds.
The near-infrared spectra display several emission lines, a feature typical of EXor-type eruptive young stars.
We estimated the mass accretion rate during the dimming phase to be $\dot{M} = 3-8 \times 10^{-7}$\,M$_\odot$\,yr$^{-1}$, higher than typical T~Tauri stars of similar mass and comparable to other EXors.
We conclude Gaia20eae is a new EXor-type candidate.
\end{abstract}


\section{Introduction} \label{sec:intro}
Optical and near-infrared photometric variability has been detected in a large number of young stellar objects (YSOs) \citep[e.g.][]{Megeath2012_AJ144192M,Bhardwaj2019_AA627A135B,Park2021_ApJ920132P}.
The most powerful cases of this variability are those of \emph{eruptive young stars} \citep{Hartmann1996_ARAA34207H,Audard2014_prplconf387A}.
These objects experience outburst events, in which their luminosity increases up to two orders of magnitude, that are detected as $2-5$\,mag brightening in optical and near-infrared bands.
These outbursts are caused by a significant increase of the mass accretion rate from $10^{-10}$--$10^{-8}$\,M$_\odot$\,yr$^{-1}$ in quiescence to $10^{-6}$--$10^{-4}$\,M$_\odot$\,yr$^{-1}$ during outburst.

Eruptive young stars are typically divided into two sub-categories: EX~Lupi-type objects (EXors) and FU~Orionis-type objects (FUors).
FUors can brighten by up to 5 magnitudes in optical and near-infrared bands, and such events can last for several decades \citep[e.g.][]{Kospal2020_ApJ889148K}, and are estimated to occur once every 10$^3$--10$^4$ years \citep{Fischer2019_ApJ872183F}.
EXors brighten by 1--4 magnitudes, and their outbursts last for a few months or up to a year, re-occurring every few years \citep[e.g.][]{JurdanaSepic2018_AA614A9J}.
The FUor category was named based on the outburst of FU~Orionis \citep{Herbig1977_ApJ217693H}, while the EXors were not clearly defined until \citet{Herbig1989_ESOC33233H} did it based on the properties of EX~Lupi.
In the last few years, the scientific community has identified some eruptive young stars that cannot be easily categorized into the two aforementioned classes.
The photometric outbursts of these objects can be as powerful as some FUors and as short as EXors, e.g., V899~Mon \citep{Ninan2015_ApJ8154N,park2021_v899mon}.
The spectra of these objects can show a combination of FUor-type and EXor-type spectral features, e.g., Gaia19ajj \citep{Hillenbrand2019_AJ158240H}, and in some cases, spectral features can appear or disappear, e.g., V1647~Ori \citep[and references therein]{Connelley2018_ApJ861145C}.

EX~Lupi has shown several outbursts in the last few decades \citep[and references therein]{Herbig2007_AJ1332679H}, including its largest confirmed outburst in 2008 \citep{Jones2008_CBET12171J} when it exhibited a $\sim$5\,mag optical brightening \citep{Abraham2009_Natur459224A}.
Following this prototype, the optical spectra of EXors in quiescence resemble that of K- or early M-type dwarfs plus a weak T~Tauri-like emission spectrum \citep{Herbig2008_AJ135637H}.
The spectral energy distributions (SEDs) of these objects during outbursts can be fitted by their quiescent SED and an additional black-body object with 1000 $\lesssim$ T $\lesssim$ 4500~K \citep{Lorenzetti2012_ApJ749188L}.

\citet{Audard2014_prplconf387A} presented a list of known eruptive young stars as of 2014, which included 15 EXors, including candidates.
In the last few years, additional eruptive young stars have been detected thanks to the all sky survey carried out by the Gaia astrometric space telescope and its Gaia Photometric Science Alerts program\footnote{\url{http://gsaweb.ast.cam.ac.uk/alerts/home}}, which publishes alerts when an object experiences a significant variation of its Gaia photometry \citep{Hodgkin2021}.
Among the recent discoveries are the new EXor ESO-H$\upalpha$~99 \citep[Gaia18dvz;][]{Hodapp2019_AJ158241H}, two new FUors, i.e. Gaia17bpi \citep{Hillenbrand2018_ApJ869146H} and Gaia18dvy \citep{SzegediElek2020_ApJ899130S}, and two eruptive young stars which do not match either of the main categories, i.e. Gaia19ajj \citep{Hillenbrand2019_AJ158240H} and Gaia19bey \citep{Hodapp2020_AJ160164H}.

Here we present our analysis of \target{}.
This object triggered a Gaia alert on 2020 August 26 after brightening by 4.6\,mag in Gaia's G photometric band.
Before the Gaia alert was announced, \citet{Hankins2020_ATel139021H} reported the sudden brightening of this object in the $J$ band as observed with the Palomar Gattini-IR survey.
The shape of its light curve suggested this object as a promising candidate for a new eruptive young star.
Therefore, we carried out multiple follow-up photometric and spectroscopic observations to verify its nature as an eruptive young star and to learn about its outburst.

The structure of the paper is as follows.
In \autoref{sec:gaia20eae} we introduce the object and its surroundings.
In \autoref{sec:observations} we present our follow up photometric and spectroscopic observations, and the auxiliary photometry used in our study of the light curve and the evolution of the SED.
The results of our observations, our analysis of the collected data is presented in \autoref{sec:results_analysis}.
In \autoref{sec:discussion} we discuss the results of our analysis and the nature of the outburst.
Finally, in \autoref{sec:conclusions} we summarize our work.

\section{Location of Gaia20eae}
\label{sec:gaia20eae}

Based on its position, 19:25:40.62 +15:07:46.56 (J2000), \target{} is seen towards the Aquila constellation, close to the Galactic plane ($b=0.5^{\circ}$).
Based on Gaia's EDR3, its distance is 2.83$_{-0.62}^{+0.96}$\,kpc \citep{BailerJones2021_AJ161147B}.

\begin{figure}
\centering
\includegraphics[width=\linewidth]{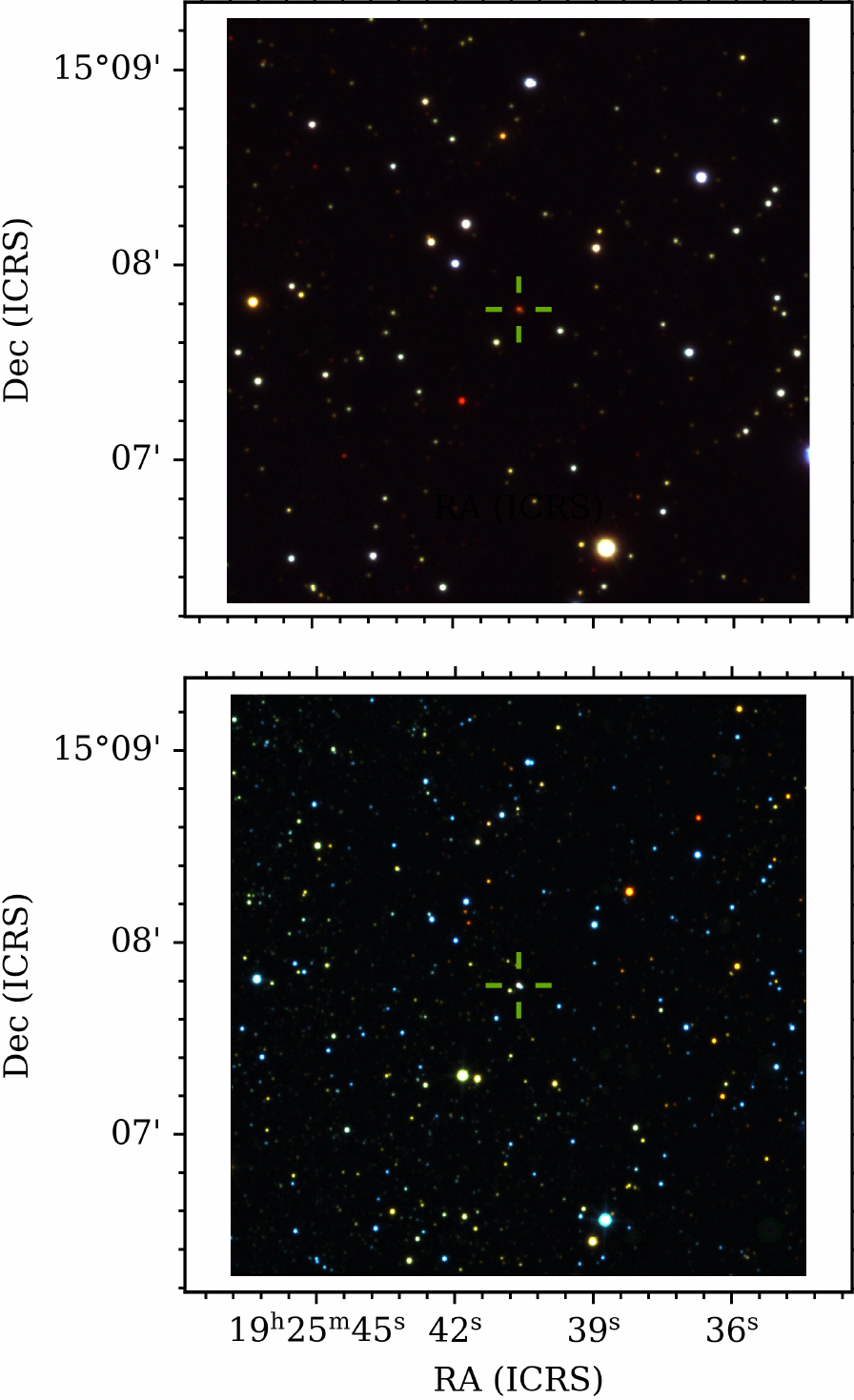}
\caption{2$'\times$2$'$ RGB composite images of \target{} (indicated by the green crosses). Top panel is Pan-STARRS $giy$ filters and bottom panel is UKIDSS $JHK_s$ filters.
\label{fig:composite_figures}}
\end{figure}

In \autoref{fig:composite_figures} we present multiple optical and infrared images of \target{} and its surroundings.
There is another {\bf likely} point source at a separation of 1$''$ towards the South-West of \target{}.
If this source is a companion of \target{}, the projected separation between them would be $\sim$2700\,au.
This faint object does not have a Gaia EDR3 parallax or proper motion so we cannot confirm whether the source is indeed part of a wide binary system with \target{}.
The possible contamination from this nearby object is analyzed in \autoref{ss:companion}.

In order to analyze whether \target{} is a member of a star forming region (SFR) we will compare its surroundings with the demographics of other well known SFRs presented by \citet{Gutermuth2009_ApJS18418G}.
First, we searched for any YSO candidates surrounding \target{} in the work of \citet{Robitaille2008_AJ1362413R}, who carried out a search for YSOs in the Galactic plane using \textit{Spitzer}/IRAC photometry.
We found that, in their catalog, an area of 2$'\times$2$'$ surrounding \target{} includes our target and an additional 18 YSO candidates.
Besides \target{}, three of these YSO candidates have EDR3 distances: SSTGLMC~G050.2675-00.5116 at 2.9$_{-0.31}^{+0.38}$\,kpc, SSTGLMC~G050.2645-00.5107 at 2.8$_{-0.31}^{+0.29}$\,kpc, and SSTGLMC~G050.2443-00.5100 at 2.1$_{-0.29}^{+0.40}$\,kpc \citep{BailerJones2021_AJ161147B}.
Two of these are the same as \target{} within their uncertainties, and the other is 0.7\,kpc closer to the Sun than \target{}.
The YSOs without distances are likely too embedded to be detected by Gaia.
If we assume that the 19 YSOs are at comparable distances, i.e.\, part of the same SFR, we obtain an average of 6.7~sources per pc$^2$.
This value is lower than almost every SFR analyzed by \citet{Gutermuth2009_ApJS18418G}.
However, our value should be considered as a lower limit because of the sensitivity of the \textit{Spitzer} data used by \citet{Robitaille2008_AJ1362413R}.
They reported that all 1\,L$_\odot$ YSOs should be detected at a distance of 0.8--1\,kpc, and since \target{} is approximately 3 times further away, the limiting luminosity is higher by almost an order of magnitude.
Demographics of SFRs indicate that most of them are populated by sources of about 1\,L$_\odot$ \citep[e.g.][]{Wilking1989_PASP101229W,Myers2012_ApJ7529M}, which would be undetectable by \textit{Spitzer} at the distance of \target{}.
Our lower limit strengthens the suggestion that our target is a YSO, and is part of a group of young stars located at similar distances.
Finally, our suggestion of the existence of this star forming region is also supported by the MC2 molecular cloud of \citet{RetesRomero2017_ApJ839113R}.
The cloud's distance, $3.4\pm0.4$\,kpc, is in agreement with \target{} and the surrounding YSOs.
It is in the direction of our target, and has projected area of $30' \times 36'$.

\section{Observations and auxiliary photometry} \label{sec:observations}
\subsection{Optical photometry}\label{ss:opt_photometry}
We started the photometric monitoring of \target{} at the beginning of 2020 September. At the Piszk\'estet\H{o} Mountain Station of Konkoly Observatory (Hungary), we used the 80\,cm Ritchey-Chretien (RC80) telescope equipped with an FLI PL230 CCD camera, 0$\farcs$55 pixel scale, $18\farcm8\times18\farcm8$ field of view, Johnson $BV$ and Sloan $g'r'i'$ filters.
At the Mount Suhora Observatory (MSO) of the Cracow Pedagogical University (Poland), we used the 60\,cm Carl-Zeiss telescope equipped with an Apogee Aspen-47 camera, 1$\farcs$116 pixel scale, $19\farcm0\times19\farcm0$ field of view, Johnson $BVRI$ and Sloan $g'r'i'$ filters.
At Adiyaman University Observatory (Turkey), we used ADYU60, a PlaneWave 60\,cm f/6.5 corrected Dall-Kirkham Astrograph telescope, equipped with an Andor iKon-M934 camera, 0$\farcs$673 pixel scale, $11\farcm5\times11\farcm5$ field of view, and Johnson $BVRI$ and Sloan $g'r'i'$ filters. 
We corrected the images for bias, dark and flat-field in a standard way and obtained aperture photometry for Gaia20eae and 20--40 comparison stars in the field of view.
The photometric calibration was done based on the APASS9 catalog \citep{henden2015}, which provides Bessell $BV$ and Sloan $g'r'i'$ magnitudes for the potential comparison stars.
We calculated the $R_C$ and $I_C$ magnitudes of the comparison stars by plotting their broad-band SED using their APASS9 and 2MASS magnitudes \citep{cutri2003} and spline interpolating for the effective wavelengths of the $R_C$ and $I_C$ filters.
The photometry between the RC80 and the Adiyaman telescopes is in agreement.
However, the photometry from the Suhora telescope is systematically brighter by 0.1\,mag, thus we have shifted these photometric points to match the photometry from the other two telescopes.
In \autoref{fig:lightcurve} we show the RC80 $r$-band photometry, while in \autoref{fig:lightcurve_zoom} we show the photometry from all bands with the three telescopes.
We present the results of our photometric monitoring in \autoref{tab:photometry}.

\begin{deluxetable}{ccccc}
\tablecaption{Optical photometry of \target{}.\label{tab:photometry}}
\tablehead{
\colhead{JD} & \colhead{Band} & \colhead{Mag.} & \colhead{Unc.} & \colhead{Tel.}
} 
\startdata
2459104.2869525 & $B$   & 18.4666 & 0.0348 & Suhora \\
2459104.2880351 & $V$   & 16.2503 & 0.0338 & Suhora \\
2459104.2887683 & $R_c$ & 15.0768 & 0.0389 & Suhora \\
2459104.2895019 & $I_c$ & 13.6838 & 0.0424 & Suhora \\
2459104.3587478 & $B$   & 18.6163 & 0.0149 & RC80 \\
2459104.3612598 & $V$   & 16.3988 & 0.0112 & RC80 \\
2459104.3623684 & $r$   & 15.5483 & 0.0118 & RC80 \\
2459104.3631305 & $i$   & 14.3834 & 0.0103 & RC80 \\
2459105.2982267 & $V$   & 16.5346 & 0.0343 & Suhora \\
2459105.2993079 & $R_c$ & 15.3087 & 0.0399 & Suhora \\
\enddata
\tablecomments{The table is published in its entirety in the machine-readable format. A portion is shown here for guidance regarding its form and content.}
\end{deluxetable}

To support our observations, we collected optical photometry for \target{} from various sources.
We downloaded \textit{G} band photometry from the Gaia Alert Index website.
We used photometry from the Data Release 7 of the Zwicky Transient Facility \citep[ZTF;][]{Masci2019_PASP131a8003M}.
In this work, we only use \textit{r} and \textit{g} band photometry which covered the 2017 November to 2021 June time period, as there were few exposures with the \textit{i} filter.
In \autoref{fig:ztf_multi} we present four different $r$-band maps at different epochs, showing the brightness changes of \target{} with respect to its surrounding sources. We also highlight the three YSO candidates mentioned in \autoref{sec:gaia20eae} which, based on the ZTF light curves, all sustain constant brightness levels.
We gathered the photometry for these three targets to construct their SEDs.
Two of the objects (SSTGLMC G050.2645-00.5107 and SSTGLMC G050.2443-00.5100, i.e.\ labels a and c in \autoref{fig:ztf_multi}, respectively) show almost negligible infrared excess, and its $<$1\,$\upmu$m photometry appears to be photospheric.
The other object, SSTGLMC G050.2675-00.5116, has an SED similar to that of \target{} albeit with higher photospheric flux, and it shows significant differences between its 2MASS and UKIDSS photometry, which could be an indication of the object being brighter in the past.
However, as we do not have spectroscopic observations of any of these objects, we cannot discern whether they are experiencing or have experienced an accretion outburst.
The light curve was supplemented with photometry from the second data release (DR2) of the Pan-STARRS1 (PS1) survey \citep{Chambers2016_arXiv161205560C} using the Panoramic Survey Telescope and Rapid Response System \citep[Pan-STARRS;][]{Kaiser2010_SPIE.7733E..0EK}.
We extracted the DR2 $grizy$ photometry from the PS1 data home page\footnote{\url{https://Pan-STARRS.stsci.edu/}}.
As can be seen from \autoref{fig:lightcurve}, not all filters were used during all epochs in which \target{} was observed.

\subsection{Optical spectroscopy}\label{ss:opt_spec}
We obtained optical spectra of \target{} in two observatories: the first was obtained during the brightness peak, on 2020 September 8/9 using the 2\,m Liverpool Telescope \citep{Steele2004_SPIE5489679S} equipped with SPRAT (ID: XOL20B01, PI: P. Zielinski).
SPRAT is providing low-resolution (R\,=\,350) spectra in the 4020\,$\text{\AA}$ to 7994\,$\text{\AA}$ range \citep{Piascik2014_SPIE9147E8HP}.
The spectrum was reduced and normalised to absolute flux units by means of the dedicated SPRAT pipeline.

The second spectrum was obtained during the brightness decline stage, on 2021 April 9/10 using the Gran Telescopio Canarias (GTC) equipped with MEGARA (ID: GTCMULTIPLE2D-21A, PI: D. García).
MEGARA is an integral field spectrograph, sampling the sky with $0\farcs6$ resolution \citep{GildePaz2016_SPIE9908E1KG}.
We used the VPH~665-HR grating which covers a wavelength range from 6405.61\,$\text{\AA}$ to 6797.14\,$\text{\AA}$ with a spectral resolution of R=20\,050.
We reduced the spectrum using the MEGARA Data Reduction Pipeline version 0.11 \citep{Cardiel2018_zndo2270518C,Pascual2020_zndo593647P}. 

We obtained a new optical spectrum of \target{} with the high-resolution FIbre-fed Echelle Spectrograph (FIES) instrument on the NOT on 2021 May 10 and on 2021 May 31.
On both nights we obtained two spectra with $t_\mathrm{exp}=2800$\,s each.
We used a fiber with a larger entrance aperture of $2\farcs5$, which provided a spectral resolution R = 25\,000, covering the 370--900\,nm wavelength range.
We used the spectra calibrated by the NOT team.
We did not obtain optical photometry with the NOT, thus, to calibrate the flux of the spectra, we used the ZTF $r$ and $g$ photometry.
To further remove any remaining artifacts in the spectrum, we calculated the median of the four flux-scaled spectra and used it for our analysis.

\subsection{Near-infrared imaging}
On 2021 May 1, we obtained $YJHK_s$ photometry of \target{} from the GTC using EMIR \citep{Balcells2000_SPIE4008797B}.
The images obtained in seven dither positions were processed using the \textsc{PyEMIR} \citep{Pascual2010_ASPC434353P} pipeline version 0.16\footnote{\url{https://pyemir.readthedocs.io/en/latest/index.html}}.
We computed aperture photometry using the DAOPHOT package \citep{stetson1987} within the {\sc IDL} software.
As the target and the comparison stars differ in brightness, the aperture size was always adjusted for every star both to include all light, and to minimize the photometric scatter.
Finally, for the photometric calibration we used 14 nearby stars listed in the 2MASS catalogue and giving consistent results. The measured magnitudes are $Y = 14.87 \pm 0.05$, $J = 13.78 \pm 0.01$, $H = 12.31 \pm 0.02$, and $K_s = 11.09 \pm 0.07$. 

We also utilized two NOT target acquisition images obtained 1.5 month before GTC, i.e., on 2021 March 16, to measure the $K_s$-band magnitude of \target{}. We carried out aperture photometry obtaining slightly brighter state with $K_s$ magnitude of $10.96 \pm 0.05$.

\begin{figure*}
\centering
\includegraphics[width=\textwidth]{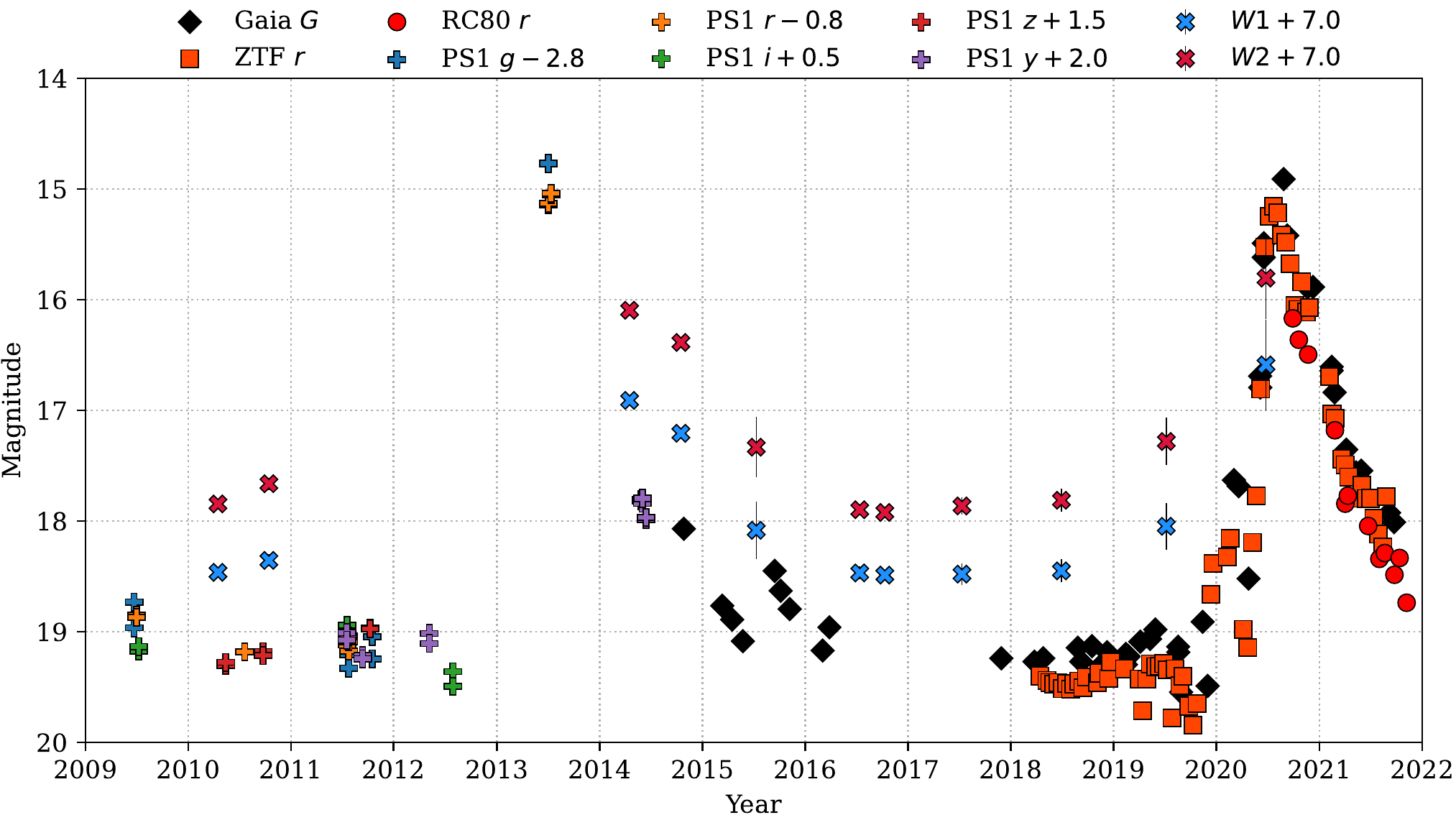}
\caption{Light curve of \target{} using optical and near-infrared (NIR) photometry.
The symbols represent the different telescopes used to construct the light curve.
The Pan-STARRS and WISE/NEOWISE photometry has been shifted as indicated by the values next to each marker label.
The ZTF and RC80 $r$ band photometry points have been binned for clarity.
Similarly, this plot only shows the \textit{r} band photometry points of our RC80 observations presented in \autoref{ss:opt_photometry}.
\label{fig:lightcurve}}
\end{figure*}

We complemented our observations with mid-infrared photometry taken by the WISE and NEOWISE missions from the NASA/IPAC Infrared Science Archive\footnote{\url{https://irsa.ipac.caltech.edu/}}.
Currently, NEOWISE observes the full sky on average twice per year with multiple exposure per epoch.
For the light curve, we computed the average and standard deviation of multiple exposures of a single epoch.
The error bars in \autoref{fig:lightcurve} are a quadratic sum of the average magnitude uncertainty per exposure.

\subsection{Near-infrared spectroscopy}\label{ss:nir_spec}
We obtained a near-infrared intermediate resolution (R=2500) $JHK_s$ spectrum of \target{} on 2021 March 17, with the $0\farcs6$ slit of the NOTCam instrument on the Nordic Observatory Telescope (ID: 61-423, PI: F. Cruz-Sáenz de Miera).
In order to correct the telluric absorption features, a telluric standard star (HD~184058, F0~V) was observed right before the target observation.
\target{} was observed with ABBA nodding pattern along the slit to subtract the sky background. 
The total exposure time in each band was 600\,s.
Flat-fielding, bad pixel removal, sky subtraction, aperture tracing, and wavelength calibration were performed for each image with \textsc{IRAF} \citep{Tody1986_SPIE627733T}.
For the wavelength calibration, an Argon lamp spectrum was used for the $J$ band and a Xenon lamp spectrum for the $H$ and $K_s$ bands.
The hydrogen absorption lines in HD~184058 were removed by Gaussian fitting.
Then, the spectrum of \target{} was divided by the normalized spectrum of HD~184058.
Finally, a barycentric velocity (20.92\,km~s$^{-1}$) correction was applied using \textsc{barycorrpy} \citep{Kanodia2018_RNAAS24K}.
We did not obtain $J$ and $H$ photometry to scale the flux of the March 17 spectra. Thus, we scaled-up the May 11 $J$ and $H$ fluxes (see below) to estimate the values for March 17. The scale factor was calculated from the ratios between the bands observed in both nights ($BVgriK_s$).

The second set of medium-resolution (R=4000--5000) spectra in $JHK_s$-bands was obtained on 2021 May 1, by means of GTC equipped with EMIR configured in the long-slit mode (PI: D. García). The star was observed through the $0\farcs6$ wide slit. The total exposure times were 1920\,s for $J$, 480\,s in $H$, and 240\,s in $K_s$ band. HgAr lamp provided wavelength calibration. The spectra were obtained in ABBA nodding pattern along the slit (exceptionally executed twice for $J$ band) and were processed by means of the dedicated PyEMIR package. The final spectrum extraction was performed under IRAF. As no telluric standards were observed during the night, we removed the telluric lines by means of {\sc molecfit} package \citep{smette2015,kausch2015}.
We calibrated the flux of these spectra using the $JHK_s$ band photometry obtained on the same night.

\section{Results \& Analysis} \label{sec:results_analysis}
\subsection{Contribution from companion}\label{ss:companion}
The companion is unresolved on all images gathered by our small telescopes and ZTF, and only marginally resolved on Pan-STARRS and EMIR-GTC images.
Thus, in the optical range we were able to estimate its contribution only during the quiescent phase based on the Pan-STARRS images obtained during the best seeing conditions: the $grizy$ data indicate on $15-25\%$ flux contribution of the fainter component to the overall \target{} brightness, and the contributions tends to be higher towards longer wavelengths.
The contribution in the infrared can only be estimated based on images obtained during the fading phase: based on EMIR observations obtained on 2021 May 1, the companion contributed to about $15-20\%$ of the total flux in $YJHK_S$ filters.
Interestingly, the contribution to the $Y$ filter is the same as that estimated from Pan-STARRS $y$-filter data during the first outburst (17\%).
Unfortunately, it is impossible to disentangle contribution of this companion to our optical spectra.
In the case of long-slit infrared EMIR spectroscopic observations, with the 0.6$''$ slit oriented N-S, contribution of this companion is almost completely removed, though this is a function of accurate pointing, guiding, and seeing variations during the individual exposures.

\subsection{Light curve}\label{ss:lightcurve}
In \autoref{fig:lightcurve} we present the light curve for \target{} composed of different photometric filters at optical and NIR wavelengths.
The photometry is based on archival data and our own observations presented in \autoref{ss:opt_photometry}.
The light curves outline two brightenings, one between 2012 and 2016, and another one that started in 2020 and triggered the Gaia alert, and which is still ongoing as of 2021, although it is now in a fading phase.
Pan-STARRS photometry indicates that the first one began between late 2012 and mid 2013, and according to Pan-STARRS \textit{y}, Gaia and NEOWISE photometry finished sometime between 2015 and 2016.
Due to the lack of photometric observations during this period, we cannot establish when this outburst reached its peak, however, based on Pan-STARRS $g$ photometry, it had an amplitude of at least 4.35\,mag with respect to the median of photometric points taken during between 2009 and 2012.
Note that the Gaia photometry in \autoref{fig:lightcurve} also shows a 0.5\,mag brightening during the second half of 2015.

\begin{figure*}
\centering
\includegraphics[width=\linewidth]{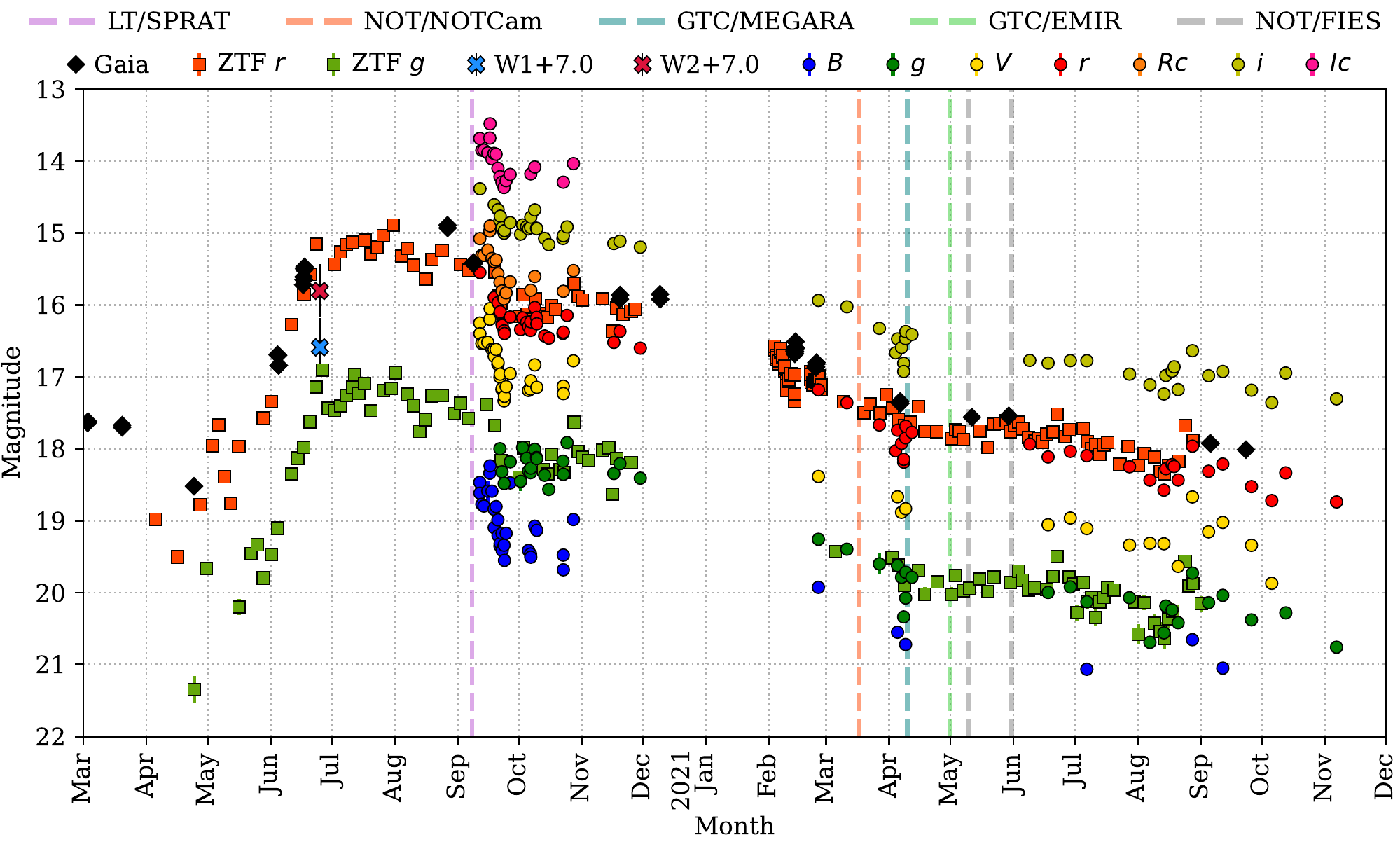}
\caption{Zoomed-in version of light curve focused on the most recent outburst in \target{} using optical photometry from multiple sources.
The symbols represent the different telescopes used to construct the light curve: diamonds are Gaia, squares are ZTF, and circles are RC80.
The vertical lines indicate the dates when we carried out our follow-up spectroscopic observations.
\label{fig:lightcurve_zoom}}
\end{figure*}

\begin{figure*}
\centering
\includegraphics[width=\textwidth]{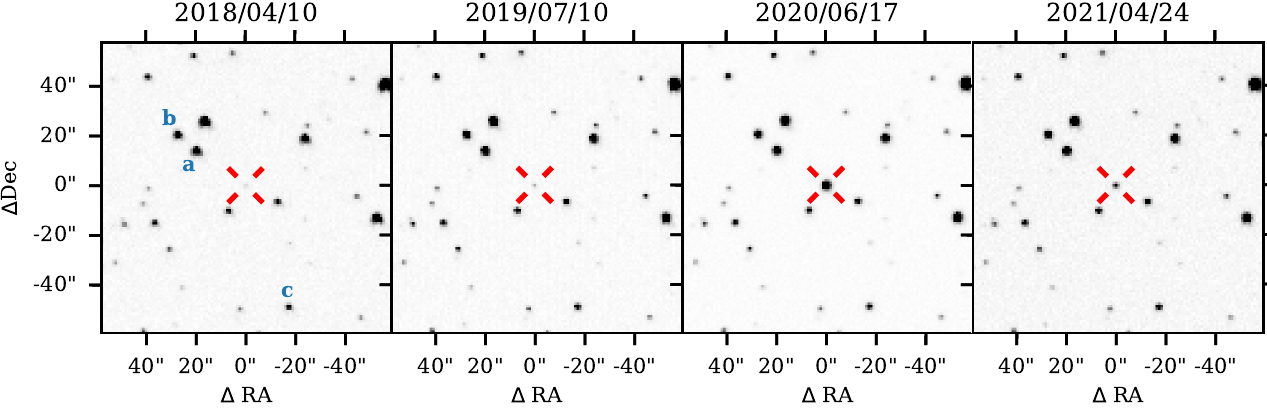}
\caption{ZTF \textit{r} band maps of \target{}, marked with the red crosshair. From left to right, the four panels represent different stages of the outburst: pre-outburst, brightening, peak and dimming. In the leftmost panel, the blue \textit{abc} labels mark the three YSO candidates with distances comparable to \target{}. These are \textit{a}: SSTGLMC~G050.2645-00.5107, \textit{b}: SSTGLMC~G050.2675-00.5116, and \textit{c}: SSTGLMC~G050.2443-00.5100.
\label{fig:ztf_multi}}
\end{figure*}

Thanks to the Gaia alert, the second brightening is better sampled with multi-band photometry.
A zoomed in light curve of this outburst in shown in \autoref{fig:lightcurve_zoom}.
The brightening began in early 2020 and, although it showed certain variability (see below), it increased with an average rate of 0.5\,mag per month between 2020 January and 2020 July.

From the NEOWISE photometry it appears this second brightening began earlier in the mid-infrared (MIR) than in the optical.
Indeed, between mid-2018 and mid-2019, \target{} was brighter in the NEOWISE bands by $\sim$0.45\,mag and in the optical bands by $\sim$0.21\,mag, and between mid-2019 and mid-2020 it was brighter by $\sim$1.45\,mag in the MIR and by $\sim$2.75\,mag in the optical.
This effect had been reported in the FUor-type outbursts of Gaia~17bpi \citep{Hillenbrand2018_ApJ869146H} and Gaia~18dvy \citep{SzegediElek2020_ApJ899130S}.

Based on ZTF photometry, the currently active brightening peaked in late 2020 July and \target{} is currently quickly fading at a rate of 0.25\,mag per month.
If this continues unchanged, it should reach its quiescence level in 2022 May.
However, this decrease in magnitude is not smooth.
\autoref{fig:lightcurve_zoom} demonstrates that starting in 2020 September the magnitude remained constant for a few months before resuming its dimming between 2020 December and 2021 January.
This light curve shows a second plateau between 2021 April and 2021 July, and it has been followed by another smooth dimming.

As mentioned earlier, there is small-scale variability throughout the light curve, thus we decided to carry out a period analysis.
A visual inspection of the ZTF $r$-band light curve shows it can be roughly separated into 5 segments.
The first is between 2018 April and 2019 February, when \target{} behaved roughly constant with small variations of 0.2\,mag.
The second is between 2019 April and 2019 November, when the light curve shows multiple 0.75\,mag dippings with its lowest being 0.87\,mag during 2019 August.
The photometry becomes sparse between 2019 November and 2020 April, however, during this third segment, \target{} is brighter than its 2018 levels by 1--1.5\,mag.
Starting in 2020 April, the frequency of the photometric points increases although it does not reach the levels of 2018--2019.
During the fourth segment, between 2020 April and 2020 July, one point shows \target{} has reached back to its quiescence brightness before starting its 4\,mag brightening.
This segment also shows a 1\,mag spike which lasted a few days.
The fifth and final segment is the dimming phase after the brightening peak.
We removed the large-scale variability of the third, fourth and fifth segments by fitting a line to the points of each segment, and subtracting it from the photometry.
For the first two segments we subtracted their respective medians.
We carried out a Lomb-Scargle analysis on each segment, however, we only found tentative evidence of a 32\,day period for the second segment.

\subsection{SED}\label{ss:qsed}
\begin{figure}
\centering
\includegraphics[width=\linewidth]{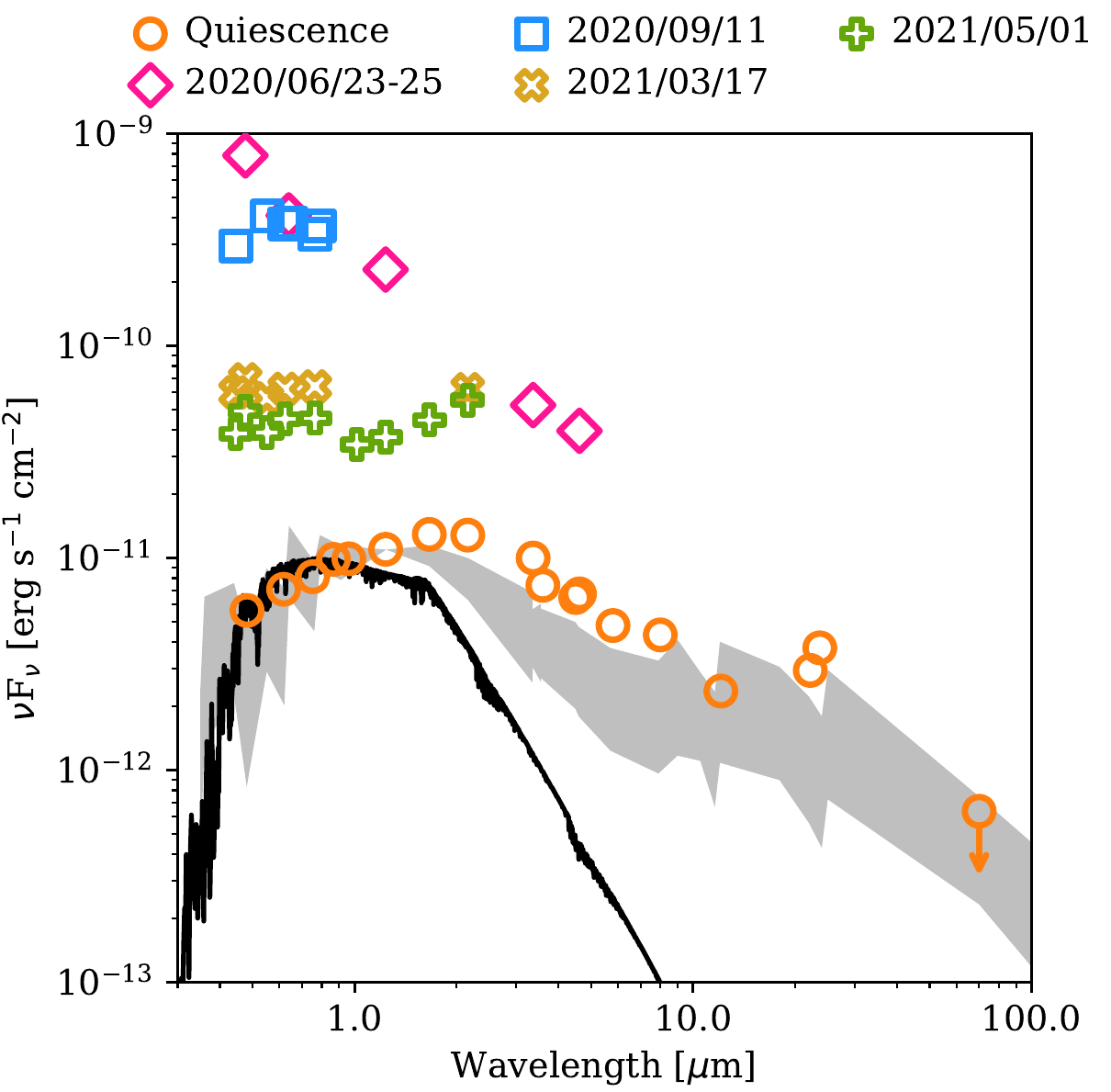}
 \caption{Dereddened pre-outburst SED constructed using the available photometry at different epochs of the light curve (see text). The pink diamonds include the $J$ photometry from the Palomar Gattini-IR survey which was taken two days before the ZTF and Gaia photometry. The black line is a NextGen model of a T$_\mathrm{eff} = 3300$\,K, $\log g = 3.5$ and solar metallicity. The shaded area represents the 25th and 75th percentiles of the SEDs of YSOs in the Taurus star forming region.\label{fig:sed}}
\end{figure}

We constructed the quiescent SED (see orange circles in \autoref{fig:sed}) using photometry from: 2MASS, WISE, Pan-STARRS (PS1), \textit{Spitzer}'s IRAC and MIPS, and \textit{Herschel}'s PACS.

We determined the spectral type and extinction in the line-of-sight to \target{} by comparing the photometric colors observed during quiescence with the intrinsic colors of pre-main sequence stars as determined by \citet{PecautMamajek13}.
The authors defined their intrinsic colors for Johnson-Cousins and 2MASS filters (see their Table 6).
For the latter, we used the 1998 2MASS photometry, and for the former, we converted the 2011 July Pan-STARRS and 1998 2MASS magnitudes to fluxes, interpolated for the wavelengths of the Johnson-Cousins filters, and computed Vega magnitudes from the fluxes.
For our best-fit we used the $V-I_C$ and $J-H$ colors as these filters are less susceptible to changes in the accretion ($B$ band) and the infrared excess caused by the disk ($K_s$ band).
We then explored the full parameter space of $T_\mathrm{eff}$ and $A_V$ (i.e. reddening the colors of certain $T_\mathrm{eff}$ by certain $A_V$) and used a $\upchi^2$ to find the best combination of parameters.
As expected, we found significant degeneracy between T$_\mathrm{eff}$ and $A_V$, however we found a best-fit at $A_V = 4.7 \pm 0.8$\,mag and $T_\mathrm{eff} = 4330 \pm 300$\,K
The bands used for the best-fit are sensitive to veiling, i.e.\ sensible to the emission of the accreting disk, and our methodology assumes photosphere-only emission.
Therefore, it is possible that we our best-fit values are currently being underestimating $T_\mathrm{eff}$ or overestimating $A_V$ due to contribution of the disk.
The best-fit values match our expectation of \target{} being a low-mass object with moderate extinction based on its location in the near-infrared color-color diagram in quiescence (see below).
The 1$\upsigma$ uncertainties were estimated from the distribution of $\upchi^2$ values.

Using the extinction-corrected pre-outburst photometric points, we estimated a quiescent bolometric luminosity of $7.22 \pm 4.0$\,L$_\odot$.
The uncertainty of our estimation is dominated by the large uncertainty of the distance estimate.
The lack of photometry at wavelengths longer than 25\,$\upmu$m means this value should be taken as a lower limit.
However, to estimate the missing bolometric luminosity, we re-calculated the luminosity with two different assumptions.
First we assumed that the value of the 70\,$\upmu$m upper limit is a detection, and second, that the real 70\,$\upmu$m flux is one third of the upper limit.
Under these scenarios, we obtained luminosities of 8.19\,L$_\odot$ and 8.16\,L$_\odot$, respectively, which are within the uncertainty of the luminosity estimated using only the shorter wavelengths.

\autoref{fig:sed} includes more sparsely sampled SEDs based on photometric observations during different points of \target{}'s light curve.
These include (i) the date of the last observations from the NEOWISE Data Release 2021 (2020 June 25), accompanied by both ZTF bands (\textit{g} and \textit{r}), (ii) the first date of our follow-up photometric monitoring (2020 September 11), (iii) the date of our NIR spectroscopic observations (2021 March 17), and (iv) the date of our GTC near infrared photometric observations (2021 May 1).
The first two of these dates show that \target{} was significantly bluer during the onset of the outburst until it started its dimming event in 2020 September.
Indeed, for 2020 June 23-25, the SED has a shape similar to a Class~III object, rather than a Class~II shape seen in the other dates.
Afterwards, the source maintains some of its blue excess, however it is comparable to a scaled-up version of the quiescent SED.
We calculated $L_\mathrm{bol}$ for these four dates and found values of 165\,L$_\odot$, 52\,L$_\odot$, 28\,L$_\odot$ and 17\,L$_\odot$, respectively.
The sparser spectral coverage of these SEDs, compared to the quiescence SED, means the underestimation factor of these values is higher.
We estimated this factor by calculating $L_\mathrm{bol}$ using the quiescence SED interpolated to the wavelengths available for each of these dates.
We found that if the quiescent SED had been as sparsely sampled as the latter dates, the quiescent bolometric luminosity would be underestimated by factors of 1.24, 6.62, 1.87, and 1.88, respectively.
The correction factor for 2020 September 11 is larger because all the photometric bands observed during that date are in the optical, consequently, the interpolation does not produce reliable results.
The corrected $L_\mathrm{bol}$ for the remaining three dates are 204\,L$_\odot$, 53\,L$_\odot$, 32\,L$_\odot$, respectively.
We would like to point out that due to the nature of the integration needed for the $L_\mathrm{bol}$ calculation, these values are highly uncertain.
However, it is worth to mention that during the 2020 outburst, the $L_\mathrm{bol}$ of \target{} appears was a factor of $\sim$28 higher than in quiescence.

\autoref{fig:sed} also includes a NextGen model based on the properties estimated earlier (black line), and the median SED of YSOs in Taurus (gray area).
At wavelengths longer than 2\,$\upmu$m, \target{} deviates from the Taurus median, this excess emission might be caused by the nearby source unresolved by the infrared facilities.

The photometric points were dereddened using the \citet{Fitzpatrick1999_PASP11163F} or the \citet{Gordon2021_ApJ91633G} extinction curves for wavelengths shorter or longer than 3\,$\upmu$m, respectively.
We used this combination of extinction curves because the former one only provides extinction values for wavelengths shorter than $\sim$3\,$\upmu$m, and the latter for wavelengths longer than 1\,$\upmu$m.
However, the extinction curves are in agreement within the overlapping wavelengths.

\subsubsection{YSO classification}
In the following, we will use different methods to check which YSO class \target{} belongs to.
We used the 2MASS H$-$K$_s$ color and the quiescent (i.e. 2010) WISE colors of \target{} (W1$-$W2\,=\,0.665, W2$-$W3\,=\,1.746, and W3$-$W4\,=\,2.236) with the color-color diagrams of \citet{Koenig2014_ApJ791131K}. 
For the W1$-$W2~versus~W2$-$W3 diagram, their Figure~5, we found that \target{} would be surrounded by Class~II objects and a few transitional disks.
In the H$-$K$_s$~versus~W1$-$W2 diagram, their Figure~6, \target{} falls among the Class~II objects of their sample.
Finally, in the W1$-$W2~versus~W3$-$W4 diagram (their Figure 7), \target{} is among Class~I, Class~II and transitional disk sources, thereby confirming our visual inspection.

The infrared spectral index $\upalpha$ is commonly used to classify YSOs into different Classes \citep{Lada1987_IAUS1151L}.
It was defined as an integral of the flux per wavelength, now it is more commonly defined as the color between two infrared bands.
We estimated the $\upalpha$ index following \citet{Kuhn2021_ApJS25433K}, who presented three different equations of calculating the spectral index based on near-infrared photometry between 4 and 24\,$\upmu$m as the interstellar extinction at these wavelengths is smaller than at shorter ones.
Fortunately, the four filters needed for the three different ways of calculating the spectral index were observed pre-outburst.
These are \textit{Spitzer}/IRAC's 4.5 and 8.0\,$\upmu$m, \textit{Spitzer}/MIPS' 24\,$\upmu$m, and WISE's W4.
We used Equations 7, 8 and 9 of \citet{Kuhn2021_ApJS25433K}, and obtained values of $\upalpha$ = $-$0.33 between the 4.5\,$\upmu$m and 24\,$\upmu$m bands, $\upalpha$ = $-$0.48 between the 4.5\,$\upmu$m and W4 bands, and $\upalpha$ = $-$0.64 between the 4.5 and 8.0\,$\upmu$m bands.
These three values fall within the range defined for Class~II objects ($-$0.3$\geq$ $\upalpha$ $\geq-$1.6), indicating \target{} is part of this Class.
Furthermore, the lack of detection at 70\,$\upmu$m (see \autoref{fig:sed}) is consistent with our classification of \target{}.

\subsubsection{Stellar radius and luminosity}\label{sss:radius_mass}
We estimated the stellar parameters of \target{} following a procedure similar to \citet{Fiorellino2021_AA650A43F}.
The best-fit effective temperature, $T_{\rm eff} = 4330$\,K, corresponds to a spectral type (SpT) of K4 \citep[][]{PecautMamajek13}.
Knowing the SpT, we can then estimate the stellar luminosity ($L_\star$) from the observed magnitudes corrected for the extinction and assuming a bolometric correction (BC). 
Then, $L_\star$ is given by:
 \begin{equation}
  \log \left( \frac{\lstar}{\lsun} \right) = 0.4[M_{\rm bol,\odot} - M_{\rm bol}]
  \label{eq:lstar}
 \end{equation}
where $M_{{\rm bol},\odot}$ is the bolometric magnitude of the Sun \citep{Mamajek15}, and the bolometric magnitude of a source is $M_{\rm bol}= m_J - 5 \log (d/10 \mbox{[pc]}) + BC_J$ where $m_J = 13.88 \pm 0.058$~mag is the extinction corrected magnitude in the quiescent phase. 
We used bolometric correction values in the $J$ band (BC$_J$) of 5--30\,Myr stars from Table~6 in \citet{PecautMamajek13}.
We found $\lstar = 4.36 \pm 0.76$~$\lsun$.

Having determined $T_\mathrm{eff}$ and $\lstar$, we used the evolutionary tracks of \citet{siess2000} to determine the mass of \target{}. 
For completeness, we point out that the most recent evolutionary tracks by \citet{baraffe15} provide masses in agreement with the models we used \citep[e.g.][]{Alcala2017_AA600A20A}.
We find $\mstar = 1.15 \pm 0.01$~$\msun$.

Based on these evolutionary models, \target{} has an age of 0.6\,Myr.
YSO timescales estimated from Spitzer photometry indicate that this age corresponds to sources with Flat spectrum or of Class~II \citep{Evans2009_ApJS181321E,Dunham2014_prplconf195D,Dunham2015_ApJS22011D}.
Recent studies have shown that approximately half of the known Flat sources have features similar to Class~II objects \citep{Heiderman2015_ApJ806231H}.
Additionally, a spectroscopic follow-up study of YSOs showed that Class~I and Flat sources show Class~II features \citep{Fiorellino2021_AA650A43F}.
Therefore, our estimated age is in agreement with the YSO Class determined earlier.

Using the same model we estimated a stellar radius of $\rstar = 3.42$~$\rsun$.
We verified this value by assuming the star emits as black-body and calculating the stellar radius with:
\begin{equation}
    \rstar = \frac{1}{2 T_{\rm eff}^2} \sqrt{\frac{\lstar}{\pi \sigma}}
    \label{rad*}
\end{equation}
where $\sigma$ is the Stefan–Boltzmann constant. 
The resulting stellar radius is $\rstar = 3.71 \pm 0.09$~$\rsun$, which is in agreement with the value provided by the model.

\subsection{Color-magnitude diagrams}
\begin{figure}
\centering
\includegraphics[width=\linewidth]{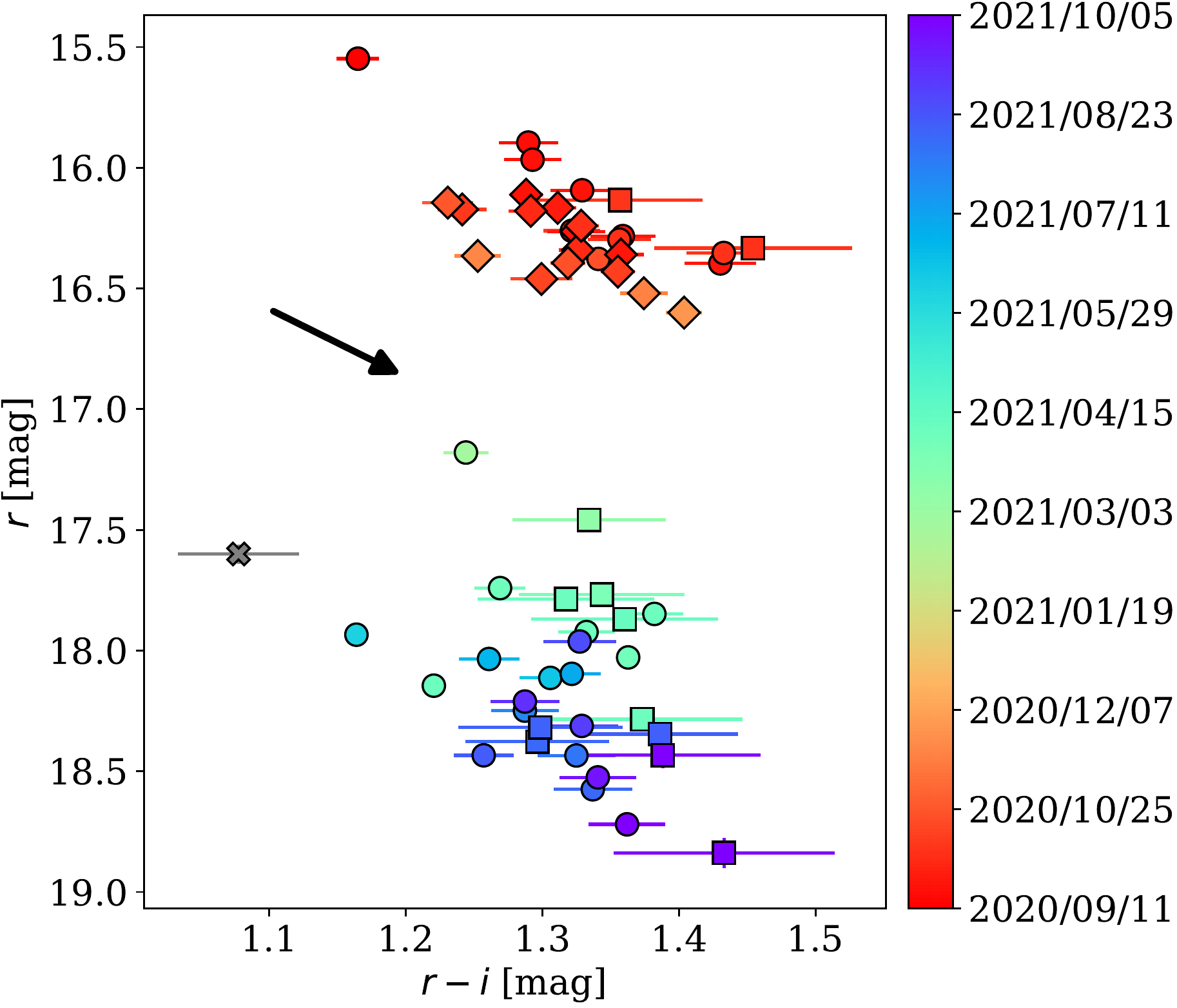}
\caption{$r$~vs.~$r-i$ color-magnitude diagram based on our follow-up photometry. The gray X symbol represents the Pan-STARRS observations 2011, an epoch of quiescence.
Circles represent data from the Konkoly RC80 telescope, the squares from the Mount Suhora telescope, and the diamonds from the Adiyaman telescope. The arrow is a sample reddening vector with $A_V=4.7$\,mag.
\label{fig:cmd_r_ri}}
\end{figure}

In \autoref{fig:cmd_r_ri} we present the $r$~versus~$r-i$ color-magnitude diagram created from the combination of the Pan-STARRS and our follow-up photometric monitoring.
For the Pan-STARRS point, we used the average of the different photometric measurements from 2011 November, a period of quiescence in \target{}, and we use this point as a reference for the pre-outburst $r-i$ color.
In the case of our follow-up photometry, we only show the points where both bands were observed on the same night, and each telescope is shown with a different symbol.
Our photometric monitoring began in early 2020 September, the month during which \target{} dimmed by 1\,mag (\autoref{fig:lightcurve_zoom}).
Following the quick dimming period of 2020 September, \target{} first began a gray fading ($r-i$ color between 1.2 and 1.4) until a brief period of variability during early 2021 April pushed the color to 1.15 before returning to its previous color value and continuing its smooth brightness decline.

\begin{figure}
\centering
\includegraphics[width=\linewidth]{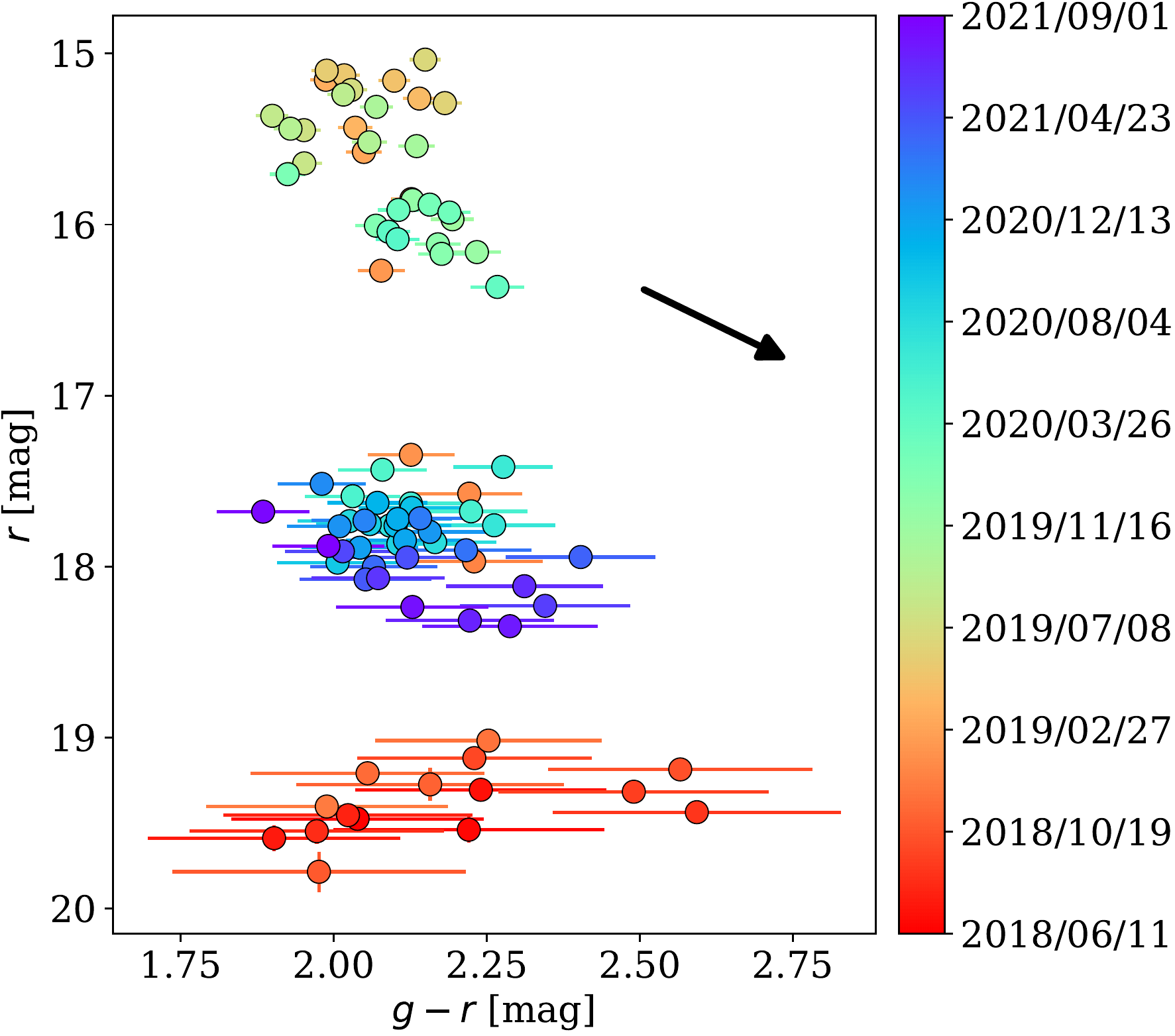}
\caption{$r$~vs.~$g-r$ color-magnitude diagram based on ZTF photometry. The arrow is a sample reddening vector with $A_V=4.7$\,mag.
\label{fig:cmd_ztf}}
\end{figure}

In \autoref{fig:cmd_ztf} we show the $r$~versus~$g-r$ color-magnitude diagram based on ZTF photometry.
Similarly to \autoref{fig:cmd_r_ri}, only the points when both bands were observed on the same night are shown.
We find the brightening and subsequent dimming of \target{} are mostly gray.
The ZTF began monitoring the sky during the quiescent period of \target{} when its brightness in $g$ was close to ZTF's median sensitivity (m$_g = 20.5$\,mag).
Therefore, the early points of the $g-r$ color are dominated by the noise of the $g$ band.

\begin{figure}
\centering
\includegraphics[width=\linewidth]{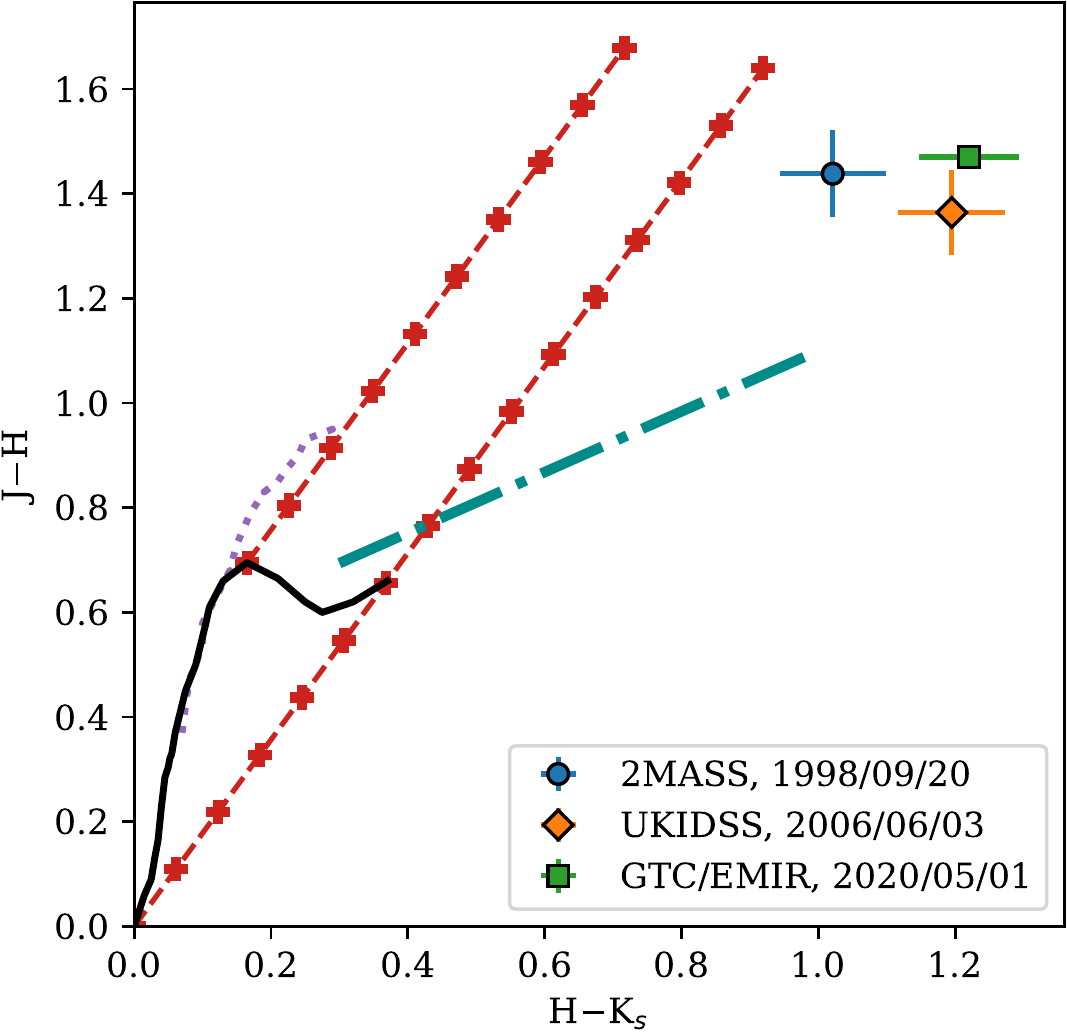}
\caption{$J-H$~vs.~$H-K_s$ color-color diagram. 
The solid black curve shows the colors of the zero-age main-sequence, and the dotted purple line represents the giant branch \citep{Bessell1988_PASP1001134B}.
The dashed red lines delimit the area occupied by the reddened normal stars with each plus sign indicating a step in $A_V$ of 1\,mag \citep{Cardelli1989_ApJ345245C}.
The dash–dotted green line is the locus of unreddened T~Tauri stars \citep{Meyer1997_AJ114288M}.  
\label{fig:ccd_jhk}}
\end{figure}

In \autoref{fig:ccd_jhk}, we present the $J-H$~versus~$H-K_s$ color-color diagram based on photometry from three different facilities at three different epochs: 2MASS on 1998 September 20, UKIDSS on 2006 June 3, and GTC on 2021 May 1.
Even though these observations were at different epochs, including one at almost the peak of outburst, we see little difference in these colors, indicating gray variability in the NIR.

\begin{figure*}
\centering
\includegraphics[width=\linewidth]{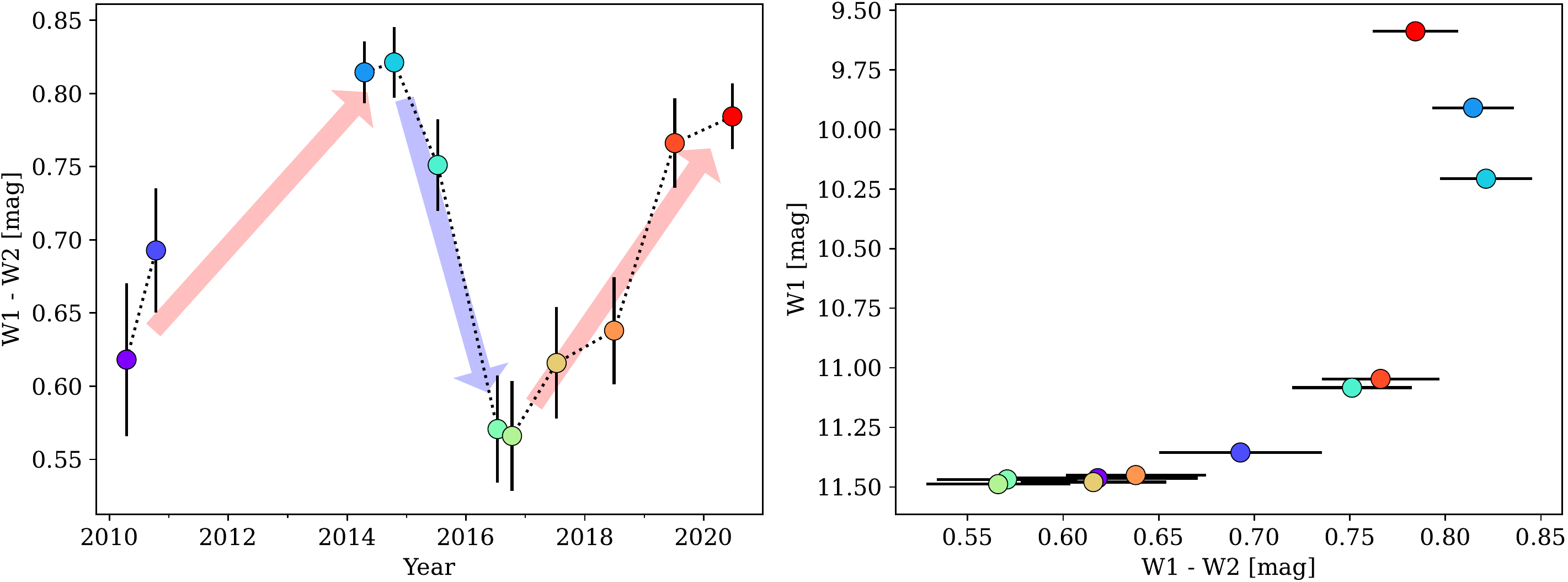}
\caption{The left panel shows the W1$-$W2 color curve of from the cryogenic WISE phase and the post-cryogenic NEOWISE phase of the project. The right panel shows the color-magnitude diagram of W1~vs.~W1$-$W2. In both panels, each combination of color and symbol represents a different date. The W1 uncertainties in the right panel are comparable to the symbol size. The panels show how \target{} became redder before the outburst reaches its peak and bluer afterwards.
\label{fig:wise}}
\end{figure*}

In \autoref{fig:wise} we show the color curve and color-magnitude diagrams constructed using the WISE and NEOWISE photometry.
The left panel shows the evolution of the W1$-$W2 color and the right panel shows the W1~versus~W1$-$W2 color-magnitude diagram.
We find \target{} is redder during the peak of its outburst and bluer during its quiescent stages.

\subsection{Optical and near-infrared spectra}
\begin{figure}
\centering
\includegraphics[width=\linewidth]{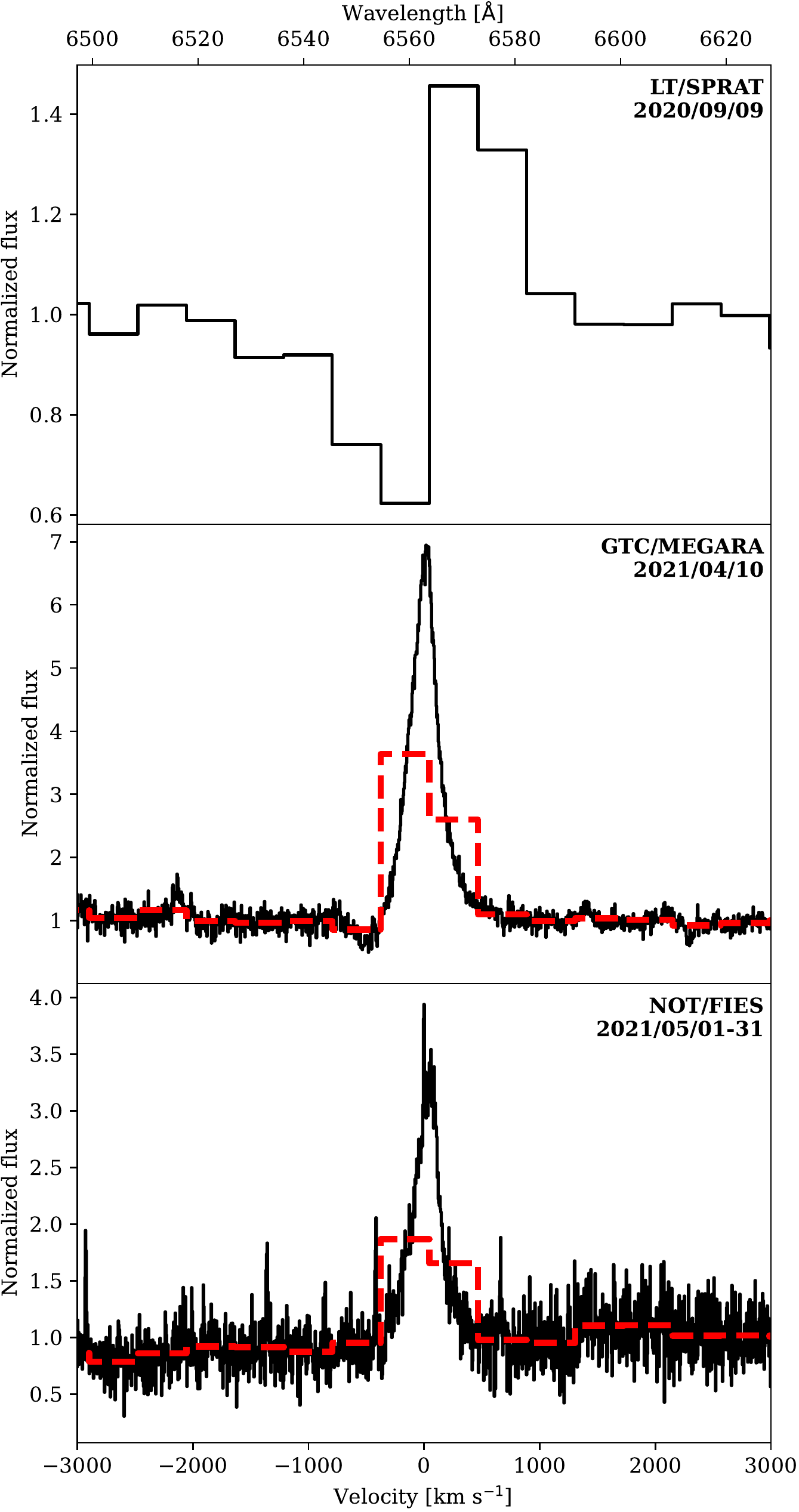}
\caption{H$\upalpha$ line profile using three different telescopes at three different dates. In the two lower panels, the red dashed line shows the spectra resampled to the spectral resolution of the Liverpool Telescope observation.\label{fig:halpha}}
\end{figure}

In \autoref{fig:halpha} we show the flux normalized line profiles of H$\upalpha$ observed at three different epochs during the latest outburst of \target{}.
For comparison purposes, we smoothed and downsampled the GTC and NOT spectra to match the Liverpool Telescope (LT) spectrum, and they are shown as red dashed lines in \autoref{fig:halpha}.
The line profiles from the LT and the GTC show a P~Cygni profile.
The LT spectrum appears to reach higher velocities than the GTC spectrum, and has a stronger blueshifted absorption.
In the case of the GTC spectrum, its redshifted emission peaks at 25\,km\,s$^{-1}$ and extends out to 450\,km\,s$^{-1}$, and its blueshifted absorption reaches its minimum at $-$530\,km\,s$^{-1}$ and extends to $-$700\,km\,s$^{-1}$.
This absorption appears to have disappeared by the time of the NOT spectrum, suggesting the blueshifted winds weakened significantly during the two to three months following the GTC observations.
We did not find significant forbidden emission lines (e.g. [\ion{Fe}{2}] and [\ion{O}{1}]).

\begin{figure*}
\centering
\includegraphics[width=\textwidth]{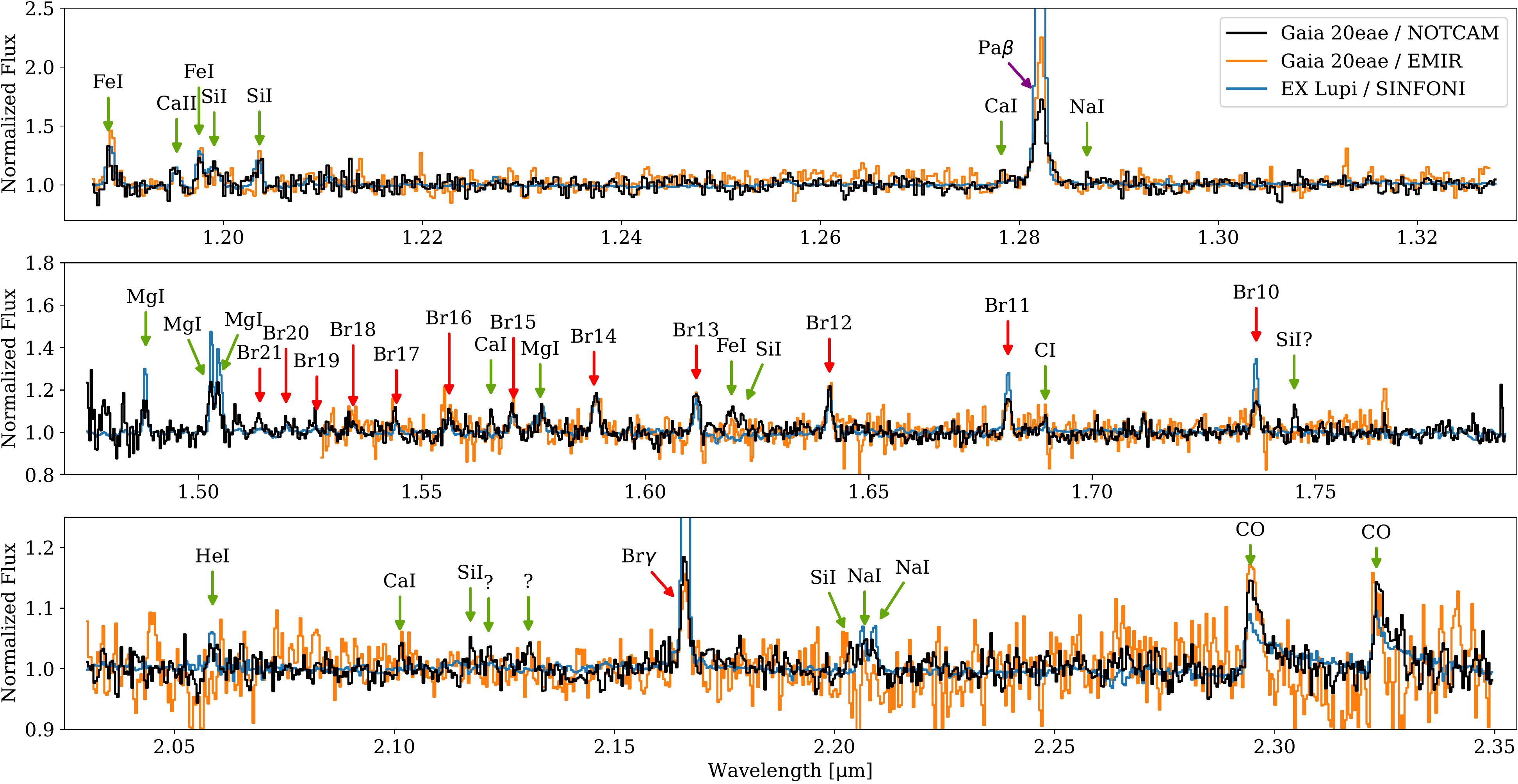}
\caption{From top to bottom, the three panels show the JHK spectra of \target{} taken with the NOT and the GTC, compared to the spectra EX~Lupi taken during its 2008 outburst with the VLT \citep{Kospal2011_ApJ7362K}. The y-axis is trimmed to showcase the weaker lines. In normalized flux peaks of Pa$\upbeta$ and Br$\gamma$ in EX~Lupi are 3.79 and 1.75, respectively. The spectra taken with EMIR and the one of EX~Lupi have been scaled to the resolution and sampling of the NOTCAM spectrum.\label{fig:spectra_jhk}}
\end{figure*}

In \autoref{fig:spectra_jhk} we present the near-IR spectra in the $JHK$ bands, and in \autoref{tab:nir_ew} we list the identified lines and their equivalent widths.
The three strongest emission lines we detected were Pa$\upbeta$, Br$\upgamma$, and the CO bandheads at 2.3\,$\upmu$m.
We also detected several emission lines from different metallic species in the three bands.
Additionally, we detected some lines from the Brackett series (Br~10 to Br~15), with some upper limits for lines up to Br~21.
To compare the properties of the \target{} spectra, we used the spectrum of EX~Lupi taken during its 2008 outburst \citep{Kospal2011_ApJ7362K}, resampled it to the spectral resolution of our NOT observations, calculated its equivalent widths, and we compared the values obtained for both young stars.
We find that in \target{}, most hydrogen lines are weaker with respect to its continuum than in the case of EX~Lupi, in particular Pa$\upbeta$ and Br$\upgamma$ which are weaker by a factor of 3.

\begin{deluxetable}{lcccc}
\tablecaption{Equivalent widths of emission lines identified in the NIR spectra of \target{} obtained with the NOT and GTC telescopes, and in the spectra of EX~Lupi obtained with the VLT \citep{Kospal2011_ApJ7362K}.\label{tab:nir_ew}}
\tablehead{
\colhead{Species} & \colhead{$\uplambda_\mathrm{obs}$} & \colhead{EW$_\mathrm{NOT}$} & \colhead{EW$_\mathrm{GTC}$} & \colhead{EW$_\mathrm{EX~Lupi}$}\\ 
 & \colhead{[$\upmu$m]} & \colhead{[$\text{\AA}$]} & \colhead{[$\text{\AA}$]} & \colhead{[$\text{\AA}$]}
} 
\startdata
\ion{Fe}{1} & 1.1884 & $-$3.11 & $-$3.67 & $-$3.02 \\
\ion{Ca}{2} & 1.1955 & $-$0.86 & $-$0.50 & $-$0.97 \\
\ion{Fe}{1} & 1.1975 & $-$2.89 & $-$1.40 & $-$3.83 \\
\ion{Si}{1} & 1.1990 & $-$3.70 & $-$3.00 & $-$4.02 \\
\ion{Si}{1} & 1.2036 & $-$2.37 & -- & $-$1.45 \\
P$\upbeta$ & 1.2820 & $-$11.19 & $-$14.76 & $-$30.27 \\
\ion{Mg}{1} & 1.4882 & $-$1.70 & -- & $-$2.82 \\
\ion{Mg}{1} & 1.5028 & $-$3.92 & -- & $-$5.44 \\
\ion{Mg}{1} & 1.5044 & $-$3.76 & -- & $-$7.06 \\
Br21 & 1.5137 & $-$1.38 & -- & $-$0.29 \\
Br20 & 1.5196 & $-$0.88 & -- & $-$0.40 \\
Br19 & 1.5265 & $-$1.04 & -- & $-$0.39 \\
Br18 & 1.5346 & $-$1.20 & $-$1.49 & $-$0.66 \\
Br17 & 1.5443 & $-$1.11 & $-$1.30 & $-$0.46 \\
Br16 & 1.5561 & $-$1.10 & $-$1.52 & $-$0.83 \\
Br15 & 1.5705 & $-$1.96 & $-$1.88 & $-$1.26 \\
\ion{Mg}{1} & 1.5768 & $-$2.02 & $-$1.17 & $-$2.22 \\
Br14 & 1.5885 & $-$3.01 & $-$3.40 & $-$2.83 \\
Br13 & 1.6114 & $-$3.25 & $-$1.16 & $-$2.03 \\
\ion{Fe}{1}? & 1.6196 & $-$2.88 & $-$0.15 & -- \\
Br12 & 1.6412 & $-$2.90 & $-$3.23 & $-$2.63 \\
Br11 & 1.6811 & $-$2.44 & $-$2.14 & $-$4.04 \\
Br10 & 1.7367 & $-$2.86 & $-$2.70 & $-$4.68 \\
\ion{O}{2}? & 1.7653 & $-$0.33 & $-$2.16 & -- \\
\ion{He}{1} & 2.0581 & $-$0.47 & -- & $-$0.78 \\
\ion{Si}{1} & 2.2035 & $-$0.50 & -- & -- \\
\ion{Na}{1} & 2.2068 & $-$0.69 & -- & $-$0.86 \\
\ion{Na}{1} & 2.2085 & $-$0.62 & -- & $-$1.00 \\
Br$\upgamma$ & 2.1661 & $-$3.92 & $-$3.42 & $-$12.15 \\
CO 2--0 & 2.2949 & $-$6.51 & $-$6.19 & $-$6.58\\
CO 3--1 & 2.3233 & $-$6.32 & $-$5.25 & $-$4.97\\
\enddata
\end{deluxetable}

\subsubsection{Mass accretion rates}
Our optical and near-infrared spectroscopy detected six different lines that are typical tracers of mass accretion rate: H$\upalpha$, the \ion{Ca}{2} IRT, Pa$\upbeta$, and Br$\upgamma$.
We used these emission lines to estimate the mass accretion rate.

First, we corrected the three spectra for the interstellar extinction of $A_V = 4.7$\,mag, as estimated above, using the extinction curve of \citet{Fitzpatrick1999_PASP11163F}.
We followed by fitting 1D Gaussian functions to the emission lines using a Markov chain Monte Carlo method using the \textsc{emcee} Python package \citep{ForemanMackey2013_PASP125306F}.
Our chain was constructed using 100 walkers and 10\,000 iterations and, after removing $\sim$300 steps of the burn-in phase, we used the remaining chain to calculate the line flux ($F_\mathrm{line}$), line luminosity ($L_\mathrm{line}$), accretion luminosity ($L_\mathrm{acc}$), and mass accretion rate ($\dot{M}_\mathrm{acc}$) using the following equations:

\begin{equation}
    F_\mathrm{line} = \sqrt{2\pi} A C
\end{equation}
\begin{equation}
    L_\mathrm{line} = 4 \pi d^2 F_\mathrm{line}
\end{equation}
\begin{equation}
    \log{L_\mathrm{acc}} = a \log{L_\mathrm{line}} + b 
\end{equation}
\begin{equation}
    \dot{M}_\mathrm{acc} = 1.25 \frac{L_\mathrm{acc} R_*}{\mathrm{G} M_*},
\end{equation}

\noindent where $A$ and $C$ are the obtained from the posterior chains of the 1D Gaussian function fitting, $d$ is the distance to the source, $a$ and $b$ are parameters empirically obtained by \citet{Alcala2017_AA600A20A} which are different for each line, G is the gravitational constant, and $R_*$ and $M_*$ are the stellar radius and mass, respectively.
For these two latter parameters, we used the values determined in \autoref{sss:radius_mass}.
The results of our calculations are presented in \autoref{tab:line_acc}.
We find that, within their uncertainties, our calculated mass accretion rates are comparable for the different lines and epochs.
The average of our estimated mass accretion rates is $(5.58 \pm 1.24) \times 10^{-7}$\,M$_\odot$\,yr$^{-1}$.

{\renewcommand{\arraystretch}{1.2}
\begin{deluxetable*}{lcccc}
\tablecaption{Emission line parameters obtained from Gaussian fitting.\label{tab:line_acc}}
\tablehead{
\colhead{Line} & \colhead{Line flux} & \colhead{Line luminosity} & \colhead{Accretion luminosity} & \colhead{Accretion rate}\\ 
 & \colhead{[erg s$^{-1}$ cm$^{-2}$]} & \colhead{[L$_\odot$]} & \colhead{[L$_\odot$]} & \colhead{[M$_\odot$ yr$^{-1}$]}
} 
\startdata
Pa$\upbeta$\tablenotemark{a} & $\left(4.44_{-0.29}^{+0.31}\right)\times10^{-14}$ & $\left(1.11_{-0.54}^{+0.70}\right)\times10^{-2}$ & $4.87_{-2.46}^{+3.32}$ & $\left(6.25_{-3.16}^{+4.26}\right)\times10^{-7}$\\
Br$\upgamma$\tablenotemark{a} & $\left(6.48_{-0.62}^{+0.60}\right)\times10^{-15}$ & $\left(1.62_{-0.80}^{+1.06}\right)\times10^{-3}$ & $5.00_{-2.77}^{+4.09}$ & $\left(6.42_{-3.56}^{+5.26}\right)\times10^{-7}$\\
\tablebreak
Pa$\upbeta$\tablenotemark{b} & $\left(3.33_{-0.16}^{+0.17}\right)\times10^{-14}$ & $\left(8.45_{-3.99}^{+5.17}\right)\times10^{-3}$ & $3.65_{-1.80}^{+2.41}$ & $\left(4.69_{-2.31}^{+3.09}\right)\times10^{-7}$\\
Br$\upgamma$\tablenotemark{b} & $\left(7.10_{-0.93}^{+1.06}\right)\times10^{-15}$ & $\left(1.81_{-0.89}^{+1.20}\right)\times10^{-3}$ & $5.71_{-3.15}^{+4.77}$ & $\left(7.34_{-4.05}^{+6.13}\right)\times10^{-7}$\\
\tablebreak
H$\upalpha$\tablenotemark{c} & $\left(2.36_{-0.05}^{+0.05}\right)\times10^{-13}$ & $\left(5.91_{-2.82}^{+3.76}\right)\times10^{-2}$ & $2.25_{-1.17}^{+1.67}$ & $\left(2.89_{-1.50}^{+2.16}\right)\times10^{-7}$\\
\ion{Ca}{2} (8498\AA)\tablenotemark{c} & $\left(5.40_{-0.09}^{+0.09}\right)\times10^{-14}$ & $\left(1.35_{-0.64}^{+0.84}\right)\times10^{-2}$ & $5.59_{-2.65}^{+3.46}$ & $\left(7.18_{-3.40}^{+4.46}\right)\times10^{-7}$\\
\ion{Ca}{2} (8542\AA)\tablenotemark{c} & $\left(5.82_{-0.12}^{+0.12}\right)\times10^{-14}$ & $\left(1.46_{-0.69}^{+0.92}\right)\times10^{-2}$ & $4.45_{-2.06}^{+2.69}$ & $\left(5.71_{-2.65}^{+3.45}\right)\times10^{-7}$\\
\ion{Ca}{2} (8662\AA)\tablenotemark{c} & $\left(4.71_{-0.20}^{+0.21}\right)\times10^{-14}$ & $\left(1.18_{-0.57}^{+0.74}\right)\times10^{-2}$ & $3.22_{-1.46}^{+1.84}$ & $\left(4.14_{-1.88}^{+2.37}\right)\times10^{-7}$\\
\enddata
\tablenotetext{a}{Observed with NOT/NOTCAM on 2021 March 17.}
\tablenotetext{b}{Observed with GTC/EMIR on 2021 May 1.}
\tablenotetext{c}{Observed with NOT/FIES on 2021 May 10/31.}
\end{deluxetable*}
}

\section{Discussion} \label{sec:discussion}
The light curve, emission lines in the infrared spectra, and the calculated values of the accretion rate point towards \target{} being an EX~Lupi-type eruptive young star.
In order to classify it as an eruptive young star, we need to verify that the mass accretion rate during the period of increased brightness is indeed higher than during the low brightness phases, but pre-brightening spectra do not exist.
Additionally, we cannot rule out that the brightness changes are periodic (with a period of $\sim$7 years) and caused by an orbiting companion.
However, we have sufficient evidence to treat \target{} as an EXor candidate so the discussion that follows is focused on the comparison between \target{} and other EXor-type eruptive young stars.

The peak of the most recent outburst of \target{}  is comparable with that of the outbursts of EX~Lupi and V1118~Ori.
The light curve of EX~Lupi's 2008 outburst can be seen in Figure~1 of \citet{Abraham2019_ApJ887156A}, and the light curve of V1118 Ori's 2019 outburst can be seen in Figure~1 of \citet{Giannini2020_AA637A83G}.
The brightening rates of \target{} and V1118~Ori are comparable at $\sim$0.5\,mag per month.
However, the brightening rates of EX~Lupi are twice as fast.
After their initial brightenings, the light curves of the EXors behave differently.
During its 2008 outburst, EX~Lupi slowly dimmed before suddenly going back to its pre-outbursting brightness at a similar rate as its brightening, and during its 2019 outburst, V1118~Ori dimmed twice as fast as it brightened.
However, the \target{} outburst is still on its dimming phase and has lasted more than one year.

The 2012/2013 \target{} outburst is sparsely sampled, however, it serves as an indication of recurrence of outbursts, as expected from a EX~Lupi-type young eruptive star.
However, the frequency of such powerful outbursts is unclear.
On one hand, there was a 53 year time difference between the two powerful outbursts of EX~Lupi, with a few weaker outbursts in between \citep{Herbig1977_ApJ217693H,Herbig2001_PASP1131547H,Herbig2007_AJ1332679H} and after \citep{Abraham2019_ApJ887156A}, and, on the other, V1118~Ori has experienced multiple 3--4\,mag outbursts since the late 1980s with only one $\sim$5\,mag outburst in 2005 \citep{Giannini2020_AA637A83G}.

The optical and NIR color-magnitude and color-color diagrams show how the source evolved during its previous outburst.
The period covered by the $r$~versus~$r-i$ color magnitude diagram (\autoref{fig:cmd_r_ri}) begins months after the outburst reached its maximum, and covers a period when \target{} dimmed by $\sim$1\,mag and reddened slightly in the lapse of one month.
The $r$~versus~$g-r$ (\autoref{fig:cmd_ztf}) shows a similar slight reddening during this short period.
After this quick reddening, the source begins a smooth gray dimming with slight color variability of $\sim$0.1\,mag.
The color-color diagram based on the $JHK_s$ photometry (\autoref{fig:ccd_jhk}) includes two observations during quiescence and one in outburst.
Typically, when a young star experiences an outburst, it becomes bluer and its location in the diagram shifts towards the T~Tauri locus \citep[e.g.][]{Lorenzetti2012_ApJ749188L,Hodapp2019_AJ158241H}.
However, our outburst photometry is almost a year after the beginning of the outburst, when \target{} has almost returned to its pre-outburst brightness, and it shows the source is slightly redder.
The reddening seen in the three diagrams suggests the source is experiencing a mild extinction increase during the dimming phase of the outburst, an effect that has been seen in other eruptive young stars, e.g. the FUor-type V1057~Cyg \citep{Szabo2021_ApJ91780S}.

In the case of the optical spectroscopy, the differences in spectral resolution and sensitivity between our three observations prevent us from carrying out a quantitative comparison between the three epochs, thus we will discuss the shape of the P~Cygni profile in a qualitative manner.
The strong blueshifted absorption during 2020 September is an indication of powerful winds related to the increased accretion rate close to the peak of the outburst.
As the mass accretion rate diminishes the blueshifted absorption feature decreases (i.e., gets closer to the continuum level), indicating that the winds have weakened.
The evolution of this powerful wind is similar to what was detected for the 2008 outburst of EX~Lupi \citep{SiciliaAguilar2012_AA544A93S}.
Indeed, the minimum of the blueshifted absorption of \target{} is at a few hundred km\,s$^{-1}$, comparable to the velocities found in the EX~Lupi winds.
However, the correlation between enhanced accretion rate and the presence of winds is not so clear in some EXors.
For example, in the case of V1118~Ori, \citet{Giannini2017_ApJ839112G} found the P~Cygni profile of H$\upalpha$ disappeared for a short period of time while the EXor was in outburst.

The near-infrared spectra show several lines in emission, as is typical of EXor-type eruptive young stars (\autoref{fig:spectra_jhk}).
When comparing the NOT and the GTC spectra, we find most detected lines are of comparable strengths.
The only difference is Pa$\upbeta$ which is 1.5 times brighter in the latter spectrum.
When comparing our observations with those of EX~Lupi during its 2008 outburst, we found \target{} has a similar set of emission lines as EX~Lupi, albeit weaker.
The difference in line strength can be explained by the different stages of the outburst in which the observations were executed.
While the EX~Lupi observations were carried out around the peak of its outburst, ours were carried out during the long-tailed dimming phase which is on-going.

Our estimates of mass accretion rates (\autoref{tab:line_acc}) are up to an order of magnitude larger than what was determined for non-eruptive classical T~Tauri stars of similar mass \citep{Fiorellino2021_AA650A43F}, and are in agreement with the mass accretion rates measured in other EXors during their outburst phases \citep[and references therein]{Audard2014_prplconf387A}.
For example, EX~Lupi had an accretion rate of $2.2 \times 10^{-7}$\,M$_\odot$\,yr$^{-1}$ during its 2008 outburst \citep{Juhasz2012_ApJ744118J}, and V1118~Ori had accretion rates of $1 \times 10^{-6}$\,M$_\odot$\,yr$^{-1}$, $1.25 \times 10^{-7}$\,M$_\odot$\,yr$^{-1}$, and $9.3 \times 10^{-8}$\,M$_\odot$\,yr$^{-1}$ during its 2005, 2016, and 2019 outbursts, respectively \citep[and references therein]{Giannini2020_AA637A83G}.
This confirms our suspicion that \target{} is a young stellar object with an elevated accretion rate.

Even though the three spectra were taken at different dates and the brightness of \target{} diminished during the two months between the first and the last spectra, the estimated mass accretion rates are the same within their 1$\upsigma$ uncertainties.
In the 45 days between the observations of the two infrared spectra, \target{} experienced a $\Updelta{}J = 0.3$\,mag and $\Updelta{}K_s = 0.1$\,mag.
It is unclear whether small changes in the brightness of an EXor reflect significant changes in the accretion.
\citet{Lorenzetti2009_ApJ6931056L} compared the Pa$\upbeta$ and Pa$\upgamma$ line fluxes with the $J$ magnitude, and the Br$\upgamma$ and CO\,2--1 line fluxes with the $K_s$.
They found that some EXors have higher/lower line fluxes when their magnitudes are lower/higher, some show changes in the line flux but not in the brightness of the object, and some have constant line fluxes regardless the brightness of the EXor.
In the case of EXors with constant line fluxes and small scale variability (e.g. UZ~Tau~E and DR~Tau), the authors propose local instabilities as an explanation for the brightness changes.
We find that within uncertainties, the line fluxes of Pa$\upbeta$ and Br$\upgamma$ did not change in the 45 days between observations.
We calculated the line fluxes for CO 2--0 and 3--1, and found they have also remained constant.
Overall, in our case, our uncertainties may be too large to detect any significant changes in the line fluxes due to small scale variability.

Alternatively, it is possible that the $a$ and $b$ coefficients used to calculate the mass accretion rates are not suitable for some eruptive young stars.
\citet{Alcala2017_AA600A20A} empirically determined the coefficients from a sample of 81 quiescent Class~II and transition disk YSOs, in which the circumstellar material is accreted onto the star via accretion flows that follow the stellar magnetic field.
In the cases with enhanced accretion rates, the magnetic field may not be strong enough to direct the accreting material to the accretion flows, thus the magnetospheric accretion model may not be valid any more. \citep{Hartmann2016_ARAA54135H}.
Motivated by this incompatibility, we calculated the mass accretion rates of \target{} during the dates of the near-infrared spectroscopic observations using the $L_\mathrm{bol}$ estimated from the available photometry (see \autoref{ss:qsed}).
Because the bolometric luminosity is composed of the stellar luminosity and the accretion luminosity, we used our corrected bolometric luminosities and the quiescent bolometric luminosity to estimate the accretion luminosities, $L_\mathrm{acc}$, using Equation 2 of \citet{White2004_ApJ616998W}.
For 2021 June 25, 2021 March 17, and 2021 May 1, the $L_\mathrm{acc}$ we obtained are 125\,L$_\odot$, 28\,L$_\odot$, and 15\,L$_\odot$.
Using Equation 6 and the same stellar parameters as before, we obtained mass accretion rates of $1.6\times10^{-5}$\,M$_\odot$\,yr$^{-1}$, $3.6\times10^{-6}$\,M$_\odot$\,yr$^{-1}$, and $2.0\times10^{-6}$\,M$_\odot$\,yr$^{-1}$, respectively.
The latter two values are an order of magnitude higher than those calculated using the empirical coefficients of \citet{Alcala2017_AA600A20A}, supporting our suggestion that they are not suitable for the outburst of \target{}, and maybe not for other outbursts.
Therefore, the mass accretion rates of other eruptive young stars that have been calculated using these empirical coefficients could also be underestimated by an order of magnitude, e.g. V1118~Ori \citep{Giannini2020_AA637A83G}.
Additionally, we would like to point out that during the peak of the outburst, the estimated mass accretion rate reached a value higher than most EXors and comparable with FUors \citep[and references therein]{Audard2014_prplconf387A}.

Finally, the sample analyzed by \citet{Alcala2017_AA600A20A} is made up of sources with luminosities below $\sim$5\,L$_\odot$, thus, it is possible that their coefficients are not suitable for objects with higher luminosities, like \target{} whose quiescent bolometric luminosity is higher than any YSO in their sample.
Indeed, the authors found evidence that their empirical relationship breaks down for objects with low stellar mass or low luminosities, and so it is possible that for luminosities higher than what was covered by their sample, the empirical relationship also breaks down.
As of this writing, a similar empirical relationship for objects with higher mass accretion rates or higher luminosities has not been obtained, therefore, our estimated mass accretion rates should be considered carefully before comparing with other objects.
Once a similar study is carried out for objects with higher accretion rates (e.g.\ Class~I YSOs), the emission line-based mass accretion rates of \target{} and other eruptive young stars could be revisited.

\section{Summary and Conclusions} \label{sec:conclusions}
The light curves of \target{} show a strong brightening in 2020 and archival photometry shows a similar outburst in 2012.
There are indications of increasing activity starting from 2018 until the 2020 outburst.
Indeed, the light curve shows $\sim$1\,mag variability in the year before the powerful brightening that triggered the two photometric alerts.

The SED in quiescence of \target{} indicates it is a Class~II YSO with $L_\mathrm{bol} \geq 7.22$\,L$_\odot$.
We find it is a low-mass K4 star with $\mstar = 1.15$\,$\msun$.
During its 2020 outburst, we estimated $L_\mathrm{bol}$ to have increased to a value of $\sim$200\,L$_\odot$, increasing by a factor slightly higher than what has been found in other EXors \citep{Audard2014_prplconf387A}.

The different epochs of H$\upalpha$ show an evolution of the P~Cygni line profile typically found in EXors.
In particular, the blueshifted absorption weakens as \target{} gets dimmer.
The $JHK_s$ spectra shows several emission lines typical of EXors such as Pa$\upbeta$, Br$\upgamma$ and higher energy transitions, and the CO 2--0 and 3--1 bandheads.

We measured the mass accretion rate using optical and near-infrared lines finding values between 2 and 7~$\times$~10$^{-7}$\,M$_\odot$\,yr$^{-1}$.
These values are higher than what has been estimated for quiescent T~Tauri stars of similar mass \citep[e.g.][]{Fiorellino2021_AA650A43F}.
The mass accretion rate estimated from the optical and near-infrared lines during a three month window shows virtually no change within the uncertainties.
However, these observations were taken during a period when the protostar had an elevated mass accretion rate, thus, the constants used to estimate the mass accretion rates might not be suitable for it.
Using the bolometric luminosities during outburst to estimate the accretion luminosities, we found that \target{} could have reach mass accretion rates as high as some FUors.

Our findings indicate that \target{} is a new EXor-type eruptive young star \emph{candidate}.
A post-outburst spectrum with lower mass accretion rates would confirm our conclusion. Another outburst detected in 2027 would indicate that the brightenings are periodic and could point towards \target{} not being an EXor.
Follow-up photometric monitoring will help to characterize the light curve and any existing periodicity.

\begin{acknowledgments}
We would like to thank the anonymous referee for the useful comments that helped us to improve the paper.
This project has received funding from the European Research Council (ERC) under the European Union’s Horizon 2020 research and innovation programme under grant agreement No 716155 (SACCRED).
We acknowledge support from the ESA PRODEX contract nr. 4000132054.
TG, BN and SA acknowledge financial support from the project PRIN-INAF 2019 ``Spectroscopically Tracing the Disk Dispersal Evolution (STRADE)'' and PRIN-INAF-MAIN-STREAM 2017 ``Protoplanetary disks seen through the eyes of new-generation instruments''.
LK acknowledges the financial support of the Hungarian National Research, Development and Innovation Office grant NKFIH PD-134784.
LK is a Bolyai János Research Fellow.
PZ and LW acknowledges the support from  European Commission's H2020 OPTICON grant No. 730890, OPTICON RadioNet Pilot grant No. 101004719 as well as Polish NCN grant Daina No. 2017/27/L/ST9/03221.
The operation of the RC80 telescope at Konkoly is supported by the project ``Transient Astrophysical Objects" GINOP 2.3.2-15-2016-00033 of the National Research, Development and Innovation Office (NKFIH), Hungary, funded by the European Union.
We acknowledge ESA Gaia, DPAC and the Photometric Science Alerts Team (\url{http://gsaweb.ast.cam.ac.uk/alerts}).
Based on the observations made with the Liverpool Telescope operated on the island of La Palma by Liverpool John Moores University in the Spanish Observatorio del Roque de los Muchachos of the Instituto de Astrofisica de Canarias with financial support from the UK Science and Technology Facilities Council. The Liverpool Telescope is operated on the island of La Palma by Liverpool John Moores University in the Spanish Observatorio del Roque de los Muchachos of the Instituto de Astrofisica de Canarias with financial support from the UK Science and Technology Facilities Council.
Based on observations made with the Gran Telescopio Canarias (GTC), installed in the Spanish Observatorio del Roque de los Muchachos of the Instituto de Astrofísica de Canarias, in the island of La Palma.
This work is (partly) based on data obtained with MEGARA instrument, funded by European Regional Development Funds (ERDF), through Programa Operativo Canarias FEDER 2014-2020.
This work is (partly) based on data obtained with the instrument EMIR, built by a Consortium led by the Instituto de Astrofísica de Canarias. EMIR was funded by GRANTECAN and the National Plan of Astronomy and Astrophysics of the Spanish Government.
\end{acknowledgments}

\begin{acknowledgments}
Based on observations obtained with the Samuel Oschin 48-inch Telescope at the Palomar Observatory as part of the Zwicky Transient Facility project.ZTF is supported by the National Science Foundation under Grant No. AST-1440341 and a collaboration including Caltech, IPAC, the Weizmann Institute for Science, the Oskar Klein Center at Stockholm University, the University of Maryland, the University of Washington, Deutsches Elektronen-Synchrotron and Humboldt University, Los Alamos National Laboratories, the TANGO Consortium of Taiwan, the University of Wisconsin at Milwaukee, and Lawrence Berkeley National Laboratories.Operations are conducted by COO, IPAC, and UW.
Based on observations obtained with the Samuel Oschin Telescope 48-inch and the 60-inch Telescope at the Palomar Observatory as part of the Zwicky Transient Facility project. ZTF is supported by the National Science Foundation under Grant No. AST-2034437 and a collaboration including Caltech, IPAC, the Weizmann Institute for Science, the Oskar Klein Center at Stockholm University, the University of Maryland, Deutsches Elektronen-Synchrotron and Humboldt University, the TANGO Consortium of Taiwan, the University of Wisconsin at Milwaukee, Trinity College Dublin, Lawrence Livermore National Laboratories, and IN2P3, France. Operations are conducted by COO, IPAC, and UW.
The Pan-STARRS1 Surveys (PS1) and the PS1 public science archive have been made possible through contributions by the Institute for Astronomy, the University of Hawaii, the Pan-STARRS Project Office, the Max-Planck Society and its participating institutes, the Max Planck Institute for Astronomy, Heidelberg and the Max Planck Institute for Extraterrestrial Physics, Garching, The Johns Hopkins University, Durham University, the University of Edinburgh, the Queen's University Belfast, the Harvard-Smithsonian Center for Astrophysics, the Las Cumbres Observatory Global Telescope Network Incorporated, the National Central University of Taiwan, the Space Telescope Science Institute, the National Aeronautics and Space Administration under Grant No. NNX08AR22G issued through the Planetary Science Division of the NASA Science Mission Directorate, the National Science Foundation Grant No. AST-1238877, the University of Maryland, Eotvos Lorand University (ELTE), the Los Alamos National Laboratory, and the Gordon and Betty Moore Foundation.
\end{acknowledgments}

\vspace{5mm}
\facilities{FLWO:2MASS, ADYU60, Gaia, GTC, Herschel, Liverpool:2m, MSO, NOT, RC80 (Konkoly), WISE, NEOWISE, ZTF}

\software{Astropy \citep{astropy},
          IRAF \citep{Tody1986_SPIE627733T}
          Matplotlib \citep{Hunter2007_matplotlib},
          MEGARA DRP \citep{Cardiel2018_zndo2270518C,Pascual2020_zndo593647P},
          PyEMIR \citep{Pascual2010_ASPC434353P},
          SciPy \citep{SciPy-NMeth2020}
         }

\bibliography{gaia20eae}{}
\bibliographystyle{aasjournal}

\end{document}